\documentclass[11pt]{article}
\usepackage[utf8]{inputenc}
\usepackage{amsmath,amsfonts,amssymb}
\usepackage{amsthm}
\usepackage{latexsym}
\usepackage[noadjust]{cite}
\usepackage{graphicx}
\usepackage{xcolor}
\usepackage{mathtools}
\usepackage[margin=1in]{geometry}
\usepackage{thm-restate}
\usepackage{enumitem}
\usepackage[linesnumbered,ruled]{algorithm2e}
\usepackage[colorlinks=true, allcolors=blue]{hyperref}
\usepackage[nameinlink,capitalise]{cleveref}
\hypersetup{
	citecolor={violet}
}

\newtheorem{theorem}{Theorem}[section]
\newtheorem{corollary}[theorem]{Corollary}
\newtheorem{lemma}[theorem]{Lemma}

\newtheorem{claim}[theorem]{Claim}
\theoremstyle{definition}
\newtheorem{definition}[theorem]{Definition}

\newtheorem{conjecture}[theorem]{Conjecture}

\newcommand{\F}{\mathbb{F}}

\newcommand{\E}{\mathbb{E}}

\newcommand{\Z}{\mathbb{Z}}

\title{
Optimal Parallel Basis Finding in Graphic and Related Matroids}
\date{\today}

\author{Sanjeev Khanna\thanks{School of Engineering and Applied Sciences, University of Pennsylvania, Philadelphia, PA. Supported in part by NSF award CCF-2402284 and AFOSR award FA9550-25-1-0107. Email: {\tt sanjeev@cis.upenn.edu}.} \and Aaron Putterman\thanks{School of Engineering and Applied Sciences, Harvard University, Cambridge, Massachusetts, USA. Supported in part by the Simons Investigator Awards of Madhu Sudan and Salil Vadhan, NSF Award CCF 2152413 and AFOSR award FA9550-25-1-0112. Email: \texttt{aputterman@g.harvard.edu}.} \and Junkai Song\thanks{School of Engineering and Applied Sciences, University of Pennsylvania, Philadelphia, PA. Email: \texttt{junkais@cis.upenn.edu}.}}

\begin{document}
	
	\maketitle
	
	\pagenumbering{gobble}
\begin{abstract}
We study the parallel complexity of finding a basis of a \emph{graphic} matroid under independence-oracle access.
Karp, Upfal, and Wigderson (FOCS 1985, JCSS 1988) initiated the study of this problem and established two algorithms for finding a spanning forest: one running in $O(\log m)$ rounds with $m^{\Theta(\log m)}$ queries, and another, for any $d \in \mathbb{Z}^+$, running in $O(m^{2/d})$ rounds with $\Theta(m^d)$ queries. A key open question they posed was whether one could simultaneously achieve \emph{polylogarithmic} rounds and \emph{polynomially many} queries. 
\medskip

We give a \emph{deterministic} algorithm that uses $O(\log m)$ adaptive rounds and $\mathrm{poly}(m)$ non-adaptive queries per round to return a spanning forest on $m$ edges, and complement this result with a matching $\Omega(\log m)$ lower bound for any (even randomized) algorithm with $\mathrm{poly}(m)$ queries per round. Thus, the adaptive round complexity for graphic matroids is characterized \emph{exactly}, settling this long-standing problem.
\medskip

Beyond graphs, we show that our framework also yields an $O(\log m)$-round, $\mathrm{poly}(m)$-query algorithm for any binary matroid satisfying a \emph{smooth circuit counting} property, implying, among others, an optimal $O(\log m)$-round  parallel algorithms for finding bases of \emph{cographic} matroids. Finally, we conjecture a natural strengthening of known circuit-counting bounds for the much broader class of \emph{regular} matroids and even an extension to so-called \emph{min-flow max-cut matroids}; assuming it, our algorithm achieves the same $O(\log m)$ rounds and $\mathrm{poly}(m)$ queries for all such matroids—which includes graphic and cographic matroids as special cases.
\end{abstract}

\vspace{5cm}
	
	\pagebreak

	\tableofcontents
	
	\pagebreak
	\pagenumbering{arabic}
	
	\section{Introduction}
	
	A key direction in combinatorial optimization is to understand how efficiently problems can be solved in parallel. This is often studied by analyzing the number of \emph{adaptive} rounds an algorithm requires, where in each round the algorithm performs a polynomial amount of non-adaptive computation. An early and highly influential contribution in this direction was made by Karp, Upfal, and Wigderson~\cite{KUW85, KUW88}, who studied the role of adaptivity in computing a basis of a matroid given access to an \emph{independence oracle}.
	Recall that a matroid $M = (E, I)$ consists of a \emph{ground set} $E$ (with $|E| = m$) and a family $I \subseteq 2^E$ of independent sets satisfying the following properties: 
	\begin{enumerate}
		\item $\emptyset \in I$.
		\item If $S \in I$ and $S' \subseteq S$, then $S' \in I$.
		\item If $S \in I$, $S' \in I$, and $|S'| < |S|$, then there exists $x \in S \setminus S'$ such that $S' \cup \{x\} \in I$.
	\end{enumerate}
	
	A \emph{basis} of a matroid is an independent set of maximal size. The task of finding a basis of a matroid is a  common generalization of many well-studied problems including finding a basis of a vector space and finding a spanning forest of a graph. 
	However, since the number of matroids on $m$ elements is super-exponential in $m$, it is generally infeasible to describe a matroid explicitly. Instead, access is typically provided via a suitable oracle. In the most basic setting (which we use in this paper), an algorithm can query an \emph{independence oracle} that, given any $S \subseteq E$, returns whether $S \in I$ (i.e., whether $S$ is independent).
	
	In the decades since~\cite{KUW85}, the study of parallel algorithms has expanded into several related areas. For instance, Blikstad~\cite{Bli22} and Blikstad and Tu~\cite{BT25} studied the related problem of \emph{matroid intersection} under independence-oracle (or rank-oracle) access. In this problem, one is given two matroids $M_1, M_2$ on the same ground set and tasked with finding a largest set of elements that is independent in both $M_1$ and $M_2$. This framework generalizes other optimization problems like bipartite matching, arborescences, and tree packing, and this line of research has culminated in algorithms requiring a sublinear (in $m$) number of rounds. Understanding parallel complexity has also emerged as a fundamental topic in the study of submodular function minimization~\cite{BS20, CCK21, CGJS22} and submodular function maximization~\cite{BS18a,BS18b,BRS19a,BRS19b,CQ19a,CQ19b,EN19,ENV19,FMZ19a,FMZ19b,KMZ+19,CFK19,BBS20,LLV20} where, similar to our setting, function evaluations are given via oracle access. There has also been extensive study on parallel algorithms for graph matchings~\cite{Lov79, KUW86, FGT16, ST17}, and parallel algorithms for finding maximal independent sets\footnote{Note that these are independent sets in the sense that they are a set of vertices without edges between them, not in the matroid sense.} in graphs~\cite{Lub86, GG24}.
	
	Despite significant progress on parallel algorithms for these various problems, many of the questions raised in~\cite{KUW85} remain unresolved. For instance, in the setting of general matroids, \cite{KUW85} showed that any algorithm making $\mathrm{poly}(m)$ queries per round must use at least $\Omega\left(\frac{m^{1/3}}{\log m}\right)$ adaptive rounds to find a basis. Conversely, they also showed that a basis of any matroid can be found in $O(\sqrt{m})$ rounds. Only recently have these bounds improved, with the work of Khanna, Putterman, and Song \cite{KPS25} presenting a new $\widetilde{O}(m^{7/15})$ round algorithm in this general case. Nevertheless, the true complexity of these problems is far from known, and as a step toward resolving the general case (as well as being a natural problem in its own right), \cite{KUW85} proposed focusing on the simpler setting of \emph{graphic matroids}.
	
	In a graphic matroid, the ground set $E$ corresponds to the edge set of a graph. A set $S \subseteq E$ of edges is \emph{independent} if no cycle is contained in $S$ (so independent sets are acyclic subgraphs), and a \emph{basis} of a graphic matroid is a spanning forest of the graph. Crucially, the algorithm sees only the oracle answers, and not the underlying graph (i.e., it can only query a set of edge labels, and the oracle returns whether that set contains a cycle). For this setting, \cite{KUW85} gave two incomparable algorithms for spanning forests via independence queries: one runs in $O(\log m)$ rounds with $m^{\Theta(\log m)}$ queries, and another, for any constant $d\in \mathbb{Z}^+$, runs in $O(m^{2/d})$ rounds with $\Theta(m^d)$ queries. Whether one can achieve \emph{simultaneously} polylogarithmic rounds and polynomially many queries has remained unresolved for four decades, serving as the central motivation of our work.

	\subsection{Our Contributions}
	
\paragraph{Graphic Matroids.}	
For the setting of graphic matroids, we provide a complete resolution of the complexity of finding a basis when given access to an independence oracle. 
    We first give a \emph{deterministic} algorithm with $O(\log m)$ rounds and $\mathrm{poly}(m)$ queries per round:
	
	\begin{theorem}
		 There is a {\em deterministic} parallel algorithm that, for any graphic matroid $G$ with $m$ elements, uses $O(\log(m))$-rounds, and at most $\mathrm{poly}(m)$ non-adaptive queries to an independence oracle per round which returns a spanning forest of $G$. 
	\end{theorem}
	
	This result resolves an open question posed by \cite{KUW85}: it shows that the adaptive round complexity can be made logarithmic \emph{without} blowing up the per-round query complexity to super-polynomial. 
	
	In fact, since the above theorem can be used to compute the rank of a graphic matroid in $O(\log(m))$ rounds, we can take advantage of \cite{BT25}'s black-box recipe for finding \emph{minimum} and \emph{maximum} weight bases\footnote{Note that in this setting, each element of the matroid is also given an associated weight. The weight of a basis is the sum of the weights of the constituent elements.}:
	
	\begin{corollary}
		There is a {\em deterministic} parallel algorithm that uses $O(\log(m))$-rounds, and at most $\mathrm{poly}(m)$ non-adaptive queries, which for any weighted, graphic matroid $G$, returns a minimum (or maximum) weight spanning forest.
	\end{corollary}
	
	We complement our algorithmic results with a matching lower bound, ruling out even a randomized algorithm with asymptotically smaller number of rounds when queries per round are restricted to be polynomial.
	
	\begin{theorem}\label{thm:lowerboundIntro}
		Let $A$ be any randomized algorithm which, for graphic matroids on $m$ elements, uses at most $\mathrm{poly}(m)$ queries to an independence oracle per round. Then, in expectation, $A$ must use $\Omega(\log(m))$ adaptive rounds to find a basis of a graphic matroid. 
	\end{theorem}
	
	Together, these results give a complete picture of the parallel complexity landscape for bases in graphic matroids, fully settling the question raised by \cite{KUW85} in this setting. 

\paragraph{Beyond Graphic Matroids.}
Beyond the graphic case, we next show that the same cycle-structural framework extends to broader classes: namely binary ($\mathbb{F}_2$-representable) matroids that satisfy a \emph{smooth circuit counting} bound.
	
	\begin{definition}[Binary Matroid]
		A matroid $M = (E, I)$ is said to be \emph{binary} ($\F_2$ representable) if there is a map from every element $e_i \in E$, to a vector $v_i$ over $\F_2$ such that a set $S \subseteq E$ is independent if and only if the corresponding set of vectors $\{v_i: e_i \in S\}$ is linearly independent in $\F_2$. 
	\end{definition}
	
	\begin{definition}[Smooth Circuit Counting Bound]\label{def:smoothcircuitcountingbound}
		For a matroid $M= (E, I)$, a \emph{circuit} in $M$ is any minimal dependent set of elements. For a set of elements $S \subseteq E$, $M|_S = (S, I \cap 2^S)$. We say that a matroid $M = (E, I)$ satisfies a smooth circuit counting bound if for every $S \subseteq E$, if $M|_S$ has minimum circuit length $\ell$, then for any $\alpha \in \Z^+$, the number of circuits of length $\leq \alpha \cdot \ell$ in $M|_S$ is at most $m^{O(\alpha)}$.
	\end{definition}
	
	Indeed, with these definitions and by leveraging the toolkit from the graphic matroid setting, we have the following more general algorithmic statement:
	
	\begin{theorem}\label{thm:representableCountingBoundAlgIntro}
    There is a randomized $O(\log m)$-round, $\mathrm{poly}(m)$-query algorithm that, for any binary matroid $M=(E,I)$ on $m$ elements satisfying the smooth circuit counting bound, finds a basis of $M$ with high probability.
	\end{theorem}
	
	In fact, this automatically implies such a bound for the class of \emph{cographic} matroids. For a graph $G$, its cographic matroid is the matroid whose circuits are precisely the \emph{cuts} of $G$.
    Cographic matroids satisfy the smooth circuit counting bound as shown by Karger’s classic cut-counting bound~\cite{Kar93}, thus yielding the corollary below:

	\begin{corollary}
		There is an algorithm which, for any cographic matroid $M$ on $m$ elements, finds a basis of $M$ in $O(\log(m))$-rounds and $\mathrm{poly}(m)$-independence queries with high probability. 
	\end{corollary}

    On the other hand, by considering the cographic matroids defined by the family of instances used in lower bound for graphic matroids (\cref{thm:lowerboundIntro}), we can also show that these $\Omega(\log(m))$ rounds are necessary for cographic matroids, thereby also settling the round complexity of basis finding in cographic matroids.
	
	Beyond this, we can even consider the well-studied class of \emph{regular} matroids (matroids that are representable over \emph{every} field). For such matroids, the binary condition is true by definition and a weakened form of the smooth circuit counting bound is known to be true. Indeed, the work of Gurjar and Vishnoi \cite{GV21} showed that in regular matroids, the number of circuits of size $\leq \alpha \ell$ (for $\ell$ being the minimum circuit size) is at most $m^{O(\alpha^2)}$. In fact, they even showed that this same bound holds for so-called ``max-flow min-cut'' matroids, a class that generalizes beyond regular matroids. Intuitively, these are matroids where certain circuit structures in the matroid (analogous to the max flow) are equal to an appropriately defined structure relating to complements of circuits in the \emph{dual matroid} (analogous to the minimum cut). Such matroids have seen extensive study in their own right, and we refer the reader to \cite{seymour1977matroids, seymour1980decomposition, GV19} for further discussion.
    
    Nevertheless, this raises the tantalizing possibility of whether the counting bound of \cite{GV19} can be strengthened to match the asymptotics of \cref{def:smoothcircuitcountingbound}:
	
	\begin{conjecture}\label{conj:intro}
		Let $M = (E, I)$ be a max-flow min-cut matroid on $m$ elements with minimum circuit length $\ell$. Then, for any $\alpha \in \Z^+$, $M$ has at most $m^{O(\alpha)}$ circuits of length $\leq \alpha \cdot \ell$.
	\end{conjecture}
	
	As stated above, a proof of \cref{conj:intro} would immediately imply an $O(\log(m))$-round, $\mathrm{poly}(m)$-query algorithm for all max-flow min-cut matroids. Likewise a proof of \cref{conj:intro} only for the smaller class of \emph{regular matroids} would also imply an algorithm of the same complexity for regular matroids.
    With the statement of our results provided, we now proceed to a discussion of the techniques we use to prove them. 
	
	\paragraph{A High Level Discussion of Techniques and Comparison With Prior Work}
	
	Note that despite the recent progress in parallel basis finding algorithms for the general matroid case \cite{KPS25}, the techniques in this paper are entirely disjoint. At a high level, \cite{KPS25} relies on a polynomial round matroid decomposition technique, which is based on random permutations, prefix circuits, and so-called greedily-optimal sets. This decomposition then enables basis finding via contraction on large independent sets and deleting elements in the general matroid setting, and does not yield any new results for the graphic matroid setting. 
	
	Our work instead develops new cycle-structure tools specific to graphs and certain other classes of matroids. These provide bounds on cycle overlaps, introduce randomized sampling to isolate short cycles, and allow for derandomization yielding universal query sets. This avoids the need for contractions altogether and instead only iteratively deletes redundant edges (while preserving connectivity). On a more technical level,  our results crucially rely on the underlying matroid obeying smooth circuit counting bounds (see \cref{def:smoothcircuitcountingbound}).  While such bounds hold for some classes of matroids, they are \emph{not} true for all matroids (for instance, they even fail for \emph{uniform matroids}).
	
	This fundamental difference is also reflected in the results: general matroids admit only polynomial-round algorithms for basis finding, with a known $\widetilde{\Omega}(m^{1/3})$ round lower bound. In this work, we instead obtain tight logarithmic-round algorithms for graphic and similarly structured matroids. This constitutes the first progress in this specific setting, thereby settling the question of the complexity of the graphic case, as posed by \cite{KUW85}.
	
	\subsection{Technical Overview}
	
	In this section, we describe the techniques that underlie our contributions in more detail. Before explaining our exact algorithms however, we provide a brief overview of how prior algorithms work, in order to give a better context for our contribution.

	\subsubsection{Prior Work}
	
	As part of the existing algorithmic literature for finding bases of graphic matroids, we discuss two broad paradigms for making progress towards recovering bases in general matroids (and therefore also for graphic matroids):
	
	\paragraph{Deleting Redundant Elements}
	
	The first paradigm used in \cite{KUW85} for recovering bases of graphic matroids is the deletion of redundant elements. More specifically, given a graphic matroid $M$ and a subset $T$ of the elements in $M$, we can say that an element $b \in M$ is \emph{redundant given $T$} if there is a cycle in $T \cup b$ that uses the element $b$. Beyond this, one can imagine recovering an entire set $B$, where for every element $b \in B$, the element $b$ forms a cycle using elements in $T$. 
	
	Because these are graphic matroids (and therefore by definition model a graph), deleting these redundant elements \emph{does not} alter the connected components of the underlying graph. In other words, if $B$ is a set of redundant edges then the {\em sparser} graph encoded by $M\setminus B$ has the same connectivity structure as the graph encoded by $M$.
	
	\paragraph{Contracting on Independent Sets}
	
	The second paradigm used in the literature is to contract on large independent sets. Given a graphic matroid $M$, recall that a set of edges $T$ is \emph{independent} if the subgraph defined by $T$ has no cycles. \emph{Contracting} on $T$ means merging the endpoints of every edge in $T$ (akin to standard graph edge contractions~\cite{BK96}). As long as $T$ is independent, the contracted matroid $M / T$ has smaller rank: in fact, $\mathrm{rank}(M/T) = \mathrm{rank}(M) - |T|$. Moreover, independence queries to $M/T$ can be simulated by queries to $M$: any query $S \subseteq M/T$ (a set of edges in the contracted graph) corresponds to querying $T \cup S$ in the original matroid $M$. Thus, contracting on an independent set $T$ reduces the problem of finding a basis of $M$ to finding a basis of the smaller matroid $M/T$.
	
	\paragraph{Using Deletion and Contraction on Graphic Matroids}
	The work of \cite{KUW85} provides a natural approach for finding bases of graphic matroids that alternates between deleting redundant elements and contracting on independent sets. To start, \cite{KUW85} fixes a parameter $d$. This can either be thought of as a fixed constant, or $O(\log(m))$, depending on which of their results one wishes to derive. Now, in a single round, \cite{KUW85} queries the independence oracle with all subsets of $\leq d$ edges from $M$. 
	
	In a single round, this requires $O(m^{d})$ queries, but also \emph{completely} enumerates all cycles of length $\leq d$. This is because for any cycle $C$ in $M$ of length $\ell \leq d$, deleting any element $x \in C$ yields an independent set, while querying the entire set $C$ yields a dependent set. Thus, just by looking at the responses to these $O(m^{d})$ queries, the algorithm can determine all cycles of length $\leq d$. Finally, after discovering all such cycles, \cite{KUW85} shows that there is a canonical way to \emph{delete} a set of redundant edges, which yields a graphic matroid where there are \emph{no} cycles of length $\leq d$, but without altering the connected components. We denote these deleted elements by $S$.
	
	However, as performed so far, this algorithm is not repeatable, as finding the cycles of length $d+1$ will require even more queries than finding those of length $\leq d$. Thus, the key observation from \cite{KUW85} is to now find a \emph{large independent set}, conditioned on the fact that there are no cycles of length $\leq d$ in the graph. In particular, the authors show that subsampling the elements at rate approximately $\frac{1}{m^{1/d}}$ yields an independent set (set of edges with no cycles) with high probability (and can in fact be derandomized). Once such an independent set is found (denote this by $T$), the algorithm then contracts on $T$. Thus in only $2$ rounds, and using only $O(m^{d})$ queries, the authors reduce the problem of finding a basis in $M$, to the problem of finding a basis in $(M \setminus S) / T$. The key invariant is that the \emph{rank} of the matroid decreases in each round. While initially as large as $m$, after $2$ rounds, the rank of $(M \setminus S) / T$ is now bounded by $m - m^{1 - 1/d}$. After repeating this $O(m^{1/d}\log(m))$ times, the algorithm then recovers a basis of the matroid.
	
	Note that in order to get $O(\log(m))$ rounds, $d$ \emph{must} be set to be $\widetilde \Omega(\log(m))$, as otherwise the number of elements we contract on is not sufficiently large to terminate within $O(\log(m))$ rounds. 
	
	\subsubsection{Our New Algorithm for Graphic Matroids}
	
	The key intuition behind our improved algorithm is, perhaps surprisingly, to \emph{avoid} contracting on independent sets altogether. Instead, our algorithm relies entirely on deleting redundant elements. In fact, the entire algorithm can be viewed as maintaining the following two invariants (while using only $\mathrm{poly}(m)$ queries per round):
	
	\begin{enumerate}
		\item After $\ell$ rounds, there are no cycles of length $\leq 1.01^{\ell}$ in the current graph.
		\item After $\ell$ rounds, the connected components of the graph are \emph{the same} as the starting graph. 
	\end{enumerate}
	
	Once these invariants hold, it is clear that the algorithm terminates in $O(\log m)$ rounds (here $m$ is the number of edges). Indeed, after $O(\log m)$ rounds, the minimum cycle length in the graph will exceed $m$, and the connected components remain unchanged from the start. In other words, we have exactly recovered a spanning forest of the original graph.
	
	Below, we explain how we can capture these invariants, \emph{without} blowing up the query complexity. We first present our algorithm as a randomized procedure, and then discuss the intuition for how it can be derandomized. 
	
	\paragraph{Cycle Counting Bounds and Cycle Structure}
	
	To start, we will require several useful facts about the structure of cycles in graphs. First, we recall the following ``cycle counting bound'', which has appeared in many works (see, for instance, \cite{Sub95, FGT16}):
	
	\begin{theorem}\label{thm:cycleCountingIntro}
		Let $G$ be a graph with $\leq m$ edges, and let $\ell$ denote the length of the shortest cycle in $G$. Then, for any $\alpha \in \Z^+$, $G$ has at most $(2m)^{2\alpha}$ cycles of length $\leq \alpha \cdot \ell$.
	\end{theorem}
	
	Ultimately, given a graphic matroid with minimum cycle length $\ell$, our goal will be to \emph{enumerate} all cycles of length $[\ell, 1.01\ell]$. The above theorem shows that, at the very least, \emph{the number} of such cycles is polynomially bounded. However, this cycle counting bound alone is not enough for us. In order to argue that we can enumerate cycles, our goal is ultimately to understand the \emph{overlap} that cycles can have with one another, as we shall see.
	
	Towards this direction, we establish some additional structural properties of cycles:
	\begin{claim}\label{clm:introCycleOverlap}
		Let $C, C'$ be two distinct cycles in a graph. Then the symmetric difference of $C, C'$ (denoted by $C \oplus C'$) contains a cycle.
	\end{claim}
	
	We omit the formal proof, but one can see this as follows: since $C$ and $C'$ are cycles, every vertex in $C \oplus C'$ has even degree (each vertex is incident to 0, 2, or 4 edges in the union), so $C \oplus C'$ must decompose into one or more cycles. With this fact in hand, our goal is to show that for any fixed short cycle $C$, no other cycle $C'$ can overlap too much with $C$. We formalize this in the following lemma:
	
	\begin{lemma}\label{lem:introCycleOverlap}
		Let $G$ be a graph such that there are no cycles of length $\leq \ell$. Let $C$ be a cycle of length $\leq 1.01 \ell$ in $G$ and let $C'$ be any other cycle (and denote its length by $k'$). Then, $|C' \setminus C| \geq \frac{k'}{4}$.
	\end{lemma}
	
	The lemma follows from a simple case analysis: when $|C'|$ is large, the overlap is inherently large due to $|C|$ being small; and when $|C'|$ is small, we instead use \cref{clm:introCycleOverlap} to lower bound $|C \setminus C'|$.

    Together, these results show that short cycles are not only few but also well-separated, 
which is precisely what enables efficiently isolating them via random sampling as we explain next.

	\paragraph{Random Sampling to Isolate Short Cycles}
	
	Now, our key algorithmic tool will be \emph{random sampling} of the edges in our graphic matroid. Recall that given a graphic matroid $M$, with minimum cycle length $\ell$, our goal is to enumerate \emph{all cycles} of length $[\ell, 1.01\ell]$. We will show that there is a sampling rate $p$ such that \emph{for any cycle} $C$ of length $[\ell, 1.01\ell]$, $C$ will be the \emph{unique} surviving cycle with probability $\geq \frac{1}{\mathrm{poly}(m)}$ when we sample the edges from $M$ at rate $p$. By this, we mean that every edge in the cycle $C$ is chosen during sampling, and \emph{for every other} cycle $C'$, at least one edge is \emph{not} chosen during sampling. 
	
	To see why this is the case, let us re-visit \cref{lem:introCycleOverlap} and \cref{thm:cycleCountingIntro}: together, these imply the following bound, which we call the \emph{cycle overlap counting bound}:
	
	\begin{lemma}[Cycle Overlap Counting Bound]\label{lem:cycleOverlapCountingIntro}
		Let $G$ be a graph with minimum cycle length $\ell$ and $\leq m$ edges, and let $C$ be any cycle in $G$ of length $\leq 1.01\ell$. Then, for any $\alpha \in \Z^+$, the number of cycles $C' \neq C$ for which $|C' \setminus C| \leq \frac{\alpha \ell}{4}$ is at most $(2m)^{2\alpha}$.
	\end{lemma}
	
	Next, we choose our value of $p$ carefully. In particular, if we fix a cycle $C$, our goal is for $C$ to survive the sampling of elements from $M$ at rate $p$, but \emph{for no other} cycle to survive. To achieve this, we set $p$ such that 
	\[
	p^{\ell} = \frac{1}{m^{100}}.
	\]
	
	Immediately, we can observe that for a cycle $C$ of length $\in [\ell, 1.01\ell]$, $C$ will survive sampling \emph{if and only if} every edge in $C$ is selected, and thus $C$ survives sampling with probability $p^{|C|}$. Since $|C| \leq 1.01 \ell$, we get that 
	\[
	\Pr[C \text{ survives sampling }] \geq p^{1.01\ell} \geq \frac{1}{m^{101}}.
	\]
	Next, we have to bound the probability that \emph{any other cycle} $C'$ survives sampling \emph{conditioned} on $C$ surviving sampling. Note that the fact that we condition on $C$ surviving sampling is exactly why we needed counting bounds for the \emph{cycle overlap} sizes, as opposed to simply the cycle sizes themselves. 
	
	The remainder of the proof is slightly Karger-esque \cite{Kar93}: let us fix a cycle $C'$ such that $|C' \setminus C| \in \left [\frac{\alpha\ell}{8}, \frac{\alpha \ell}{4} \right ]$. Our goal is to understand 
	\[
	\Pr[C' \text{ survives sampling} | C \text{ survives sampling}].
	\]
	To do this, we can observe that conditioned on $C$ surviving sampling, $C'$ survives sampling if and only if all the edges in $C' \setminus C$ survive sampling. Further, because these edges are disjoint from the edges in $C$, their survival probability is \emph{independent} of the edges in $C$. That is to say
	\[
	\Pr[C' \text{ survives sampling} | C \text{ survives sampling}] = \Pr[C' \setminus C \text{ survives sampling}],
	\]
	which we can now evaluate to be exactly $p^{|C' \setminus C|}$, which is $\leq p^{\frac{\alpha\ell}{8}} = \left ( \frac{1}{m^{100}} \right )^{\alpha / 8} \leq \frac{1}{m^{12\alpha}}$.
	
	However, now we use the fact that by \cref{lem:cycleOverlapCountingIntro}, there are at most $(2m)^{2\alpha}$ cycles $C'$ for which $|C' \setminus C| \in \left [\frac{\alpha\ell}{8}, \frac{\alpha \ell}{4} \right ]$. In particular, we can simply take a union bound over all such cycles and see that
	\[
	\Pr\left [\exists C' \neq C: |C' \setminus C| \in \left [\frac{\alpha\ell}{8}, \frac{\alpha \ell}{4} \right ] \wedge C' \text{ survives sampling} |C \text{ survives sampling} \right] \leq \frac{(2m)^{2\alpha}}{m^{12\alpha}}. 
	\]
	Now, integrating over $\alpha$, we see that the probability that \emph{any cycle} $C' \neq C$ survives sampling conditioned on $C$ surviving sampling is bounded above by $\frac{1}{\mathrm{poly}(m)}$. In particular, conditioned on $C$ surviving, this means $C$ is the \emph{unique} surviving cycle with probability $\geq 1/2$. Thus,
	\[
	\Pr[C \text{ is the unique surviving cycle}] 
	\]
	\[
	= \Pr[C \text{ survives}] \cdot \Pr[\forall C' \neq C, C' \text{ does not survive} | C \text{ survives}] 
	\]
	\[
	\geq \Pr[C \text{ survives}] \left(1-\frac{1}{\text{poly}(m)}\right) \geq \frac{1}{2 \cdot m^{101}}.
	\]
	
	Thus, if we repeat this sampling procedure some large polynomial number of times (say, $m^{102}$), then with extremely high probability, every cycle of length $[\ell, 1.01\ell]$ is the \emph{unique} surviving cycle in some sub-sampled graph. All that remains is to find a way to identify and remove these cycles.
	
	\paragraph{Removing Short Cycles}
Sampling ensures that each short cycle is uniquely exposed in some sampled subgraph. 
We now need a procedure to identify these cycles using only independence-oracle queries.
Given a sampled edge set $E'$, we can determine whether it contains exactly one cycle as follows:

	\begin{enumerate}
		\item  Query $\mathrm{Ind}(E')$ to test whether $E'$ is dependent. 

		\item Assuming $E'$ is dependent, we next need to ensure it has a unique cycle. If $\mathrm{Ind}(E' \setminus \{x\})$ remains dependent for all $x$, $E'$ has multiple cycles; otherwise, the edges whose removal restores independence belong to the unique cycle in $E'$.
	\end{enumerate}

    Repeating over polynomially many samples in parallel recovers all short cycles. We can then invoke the deletion procedure of \cite{KUW85} to remove these cycles without changing the connectivity structure of the graph, thus ensuring both invariants discussed above.

	\paragraph{Derandomization via Universal Query Sets} 
    The randomized procedure above already yields an $O(\log m)$-round algorithm; 
to make it \emph{deterministic}, we transform the sampling process into a fixed collection of universal query sets, 
leveraging strong probabilistic guarantees for success.
 Specifically, recall that we showed that for any graphic matroid $M$ with $\leq m$ elements and shortest cycle length $\ell$, there is a random sampling procedure which for any cycle $C$ of length $\in [\ell, 1.01\ell]$, ensures that 
	\[
	\Pr[C \text{ uniquely survives sampling}] \geq \frac{1}{2\cdot m^{101}}.
	\]
	
    If we repeating this sampling procedure $m^{200}$ times (say), and denote these random samples by $B_1, \dots B_{m^{200}}$, then for a fixed $M$, and a fixed $C$ in $M$ of length $[\ell, 1.01\ell]$,
	\[
	\Pr[\exists B_i: C \text{ uniquely survives sampling in } B_i] \geq 1 - 2^{-m^{98}}.
	\]
	Importantly, we can now take a union bound (1) over all cycles $C$ in $M$ (which is only $\mathrm{poly}(m)$), and (2) over \emph{all} possible graphic matroids $M$ on $m$ elements, since there are $\leq 2^{m^2}$ possible such matroids, establishing that there exists some choice of $B_1, \dots B_{m^{200}}$ that simultaneously recovers short cycles in \emph{all possible graphic matroids} with minimum circuit length $\ell$. 
    
    In other words, we have identified a polynomial-size \emph{universal query set} for all $m$-element graphic matroids with minimum cycle length $\ell$. Since the number of edges and minimum circuit length change over time, we provide the \emph{deterministic} algorithm with a set $B^{a,\ell}_1, \dots B^{a,\ell}_{m^{200}}$ of queries for every $a \in [m], \ell \in [m]$. This gives a (non-uniform) derandomization of our basis-finding algorithm.

    \paragraph{Conceptual Framework and Generalization Beyond Graphic Matroids}

Conceptually, our algorithm departs from contraction-based methods and develops a 
\emph{cycle-isolation framework}: by iteratively isolating and removing short dependent structures, we gradually increase the minimum circuit length while preserving rank.
This framework can thus be extended to matroids that satisfy the following two properties:

    \begin{enumerate}
        \item A smooth circuit counting bound on the \emph{total number} of cycles of length $\leq \alpha \ell$ (with $\ell$ being the minimum).\label{item1intro}
        \item Closure of circuits under symmetric difference (\cref{lem:introCycleOverlap}). \label{item2intro}
    \end{enumerate}

    To generalize beyond graphic matroids, we work with \emph{circuits} instead of cycles, where circuits are the minimal dependent sets, and observe that these conditions hold for binary ($\mathbb{F}_2$-representable) matroids satisfying a smooth circuit-counting bound. This yields a unified framework that extends our results beyond graphic matroids to \emph{cographic}, and possibly to \emph{regular} or even \emph{max-flow min-cut} matroids assuming \cref{conj:intro} holds.

	\subsubsection{Matching Lower Bounds for Graphic Matroids}

We show that any (randomized) algorithm limited to $\mathrm{poly}(m)$ independence queries per round needs $\Omega(\log m)$ adaptive rounds. For the overview, assume each round permits at most $m^c$ independence queries (for a fixed constant $c$). We construct a hard instance family and give some intuition as to why it forces $\Omega(\log m)$ rounds; the formal proof is a bit delicate, and appears in \cref{sec:lowerbound}.

	\paragraph{The Lower Bound Instance}
	
	Fix the constant $c$ and set $L \coloneqq \Theta(m^{1/2})$ and $\gamma \coloneqq (1000c)^2$. 
	Build a graph $G$ containing, for each $i=0,1,\ldots,\lfloor \log_{\gamma}(\sqrt{L})\rfloor$, 
	$L$ vertex-disjoint cycles of length $\sqrt{L}\,\gamma^{i}$ (and no other edges sharing vertices with these cycles). 
	Equivalently,
	\[
	G = \bigcup_{i=0}^{\lfloor \log_{\gamma}(\sqrt{L})\rfloor}  L\cdot C_{\sqrt{L}\,\gamma^i},
	\]
	where $C_{\ell}$ denotes a cycle of length $\ell$. This gives $|E(G)|=\Theta(L^2)=\Theta(m)$; we pad with isolated edges if needed so that the total is exactly $m$.
	
	Crucially, the algorithm never sees the graph; it sees only the answers to independence queries. 
	We therefore hide the structure by assigning uniformly random labels to the edges. 
	An independence query on a label set reveals only whether a cycle is present, not which edges form it.
	
	\paragraph{Deriving the Lower Bound}
	
	Let $G$ denote a random instance from the above class, and let us consider the first round of queries that any algorithm makes, say, 
	$B_1, \dots B_{m^c}$. The key observation is that if $\mathrm{Ind}(B_i) = 0$ (that is, there is a cycle in $B_i$), then with very high probability, it is one of the cycles of \emph{shortest length} (i.e., $\sqrt{L}$) that is in $B_i$.
		
	Since the labels are uniform, a fixed query $B_i$ is effectively a random set of $\lvert B_i\rvert$ edges. 
	Let $\beta \coloneqq \lvert B_i\rvert/\lvert E(G)\rvert$ denote its sampling rate. 
	A particular cycle of length $\ell$ is fully contained in $B_i$ with probability $\beta^{\ell}$. 
	For the shortest cycles (length $\sqrt{L}$), the expected number contained in $B_i$ is $L\cdot \beta^{\sqrt{L}}$. 
	If \emph{none} of these shortest cycles is hit, then necessarily $\beta^{\sqrt{L}}\le 1/\sqrt{L}$; otherwise, a Chernoff bound would imply that one survives w.h.p. 
	But then for longer cycles (length $\gamma\sqrt{L}$),
	\[
	\beta^{\gamma\sqrt{L}} \le \left(\tfrac{1}{\sqrt{L}}\right)^{\gamma} \le \tfrac{1}{L^{1000c}},
	\]
	so even across $m^c$ queries, the chance to capture such a long cycle is negligible.
	
	Hence, in the first round either (i) no cycle is captured, or (ii) with high probability at least one captured cycle has length $\sqrt{L}$.
	
	In summary, first-round answers are driven almost entirely by the shortest cycles:
	\[
	\mathrm{Ind}(B) = \mathbf{1}\big[\exists\ \text{cycle of length }\sqrt{L}\subseteq B\big]\quad\text{w.h.p.}
	\]
	Thus the algorithm’s information gain is essentially the identity of edges in the shortest cycles. 
	(For the argument, we will even \emph{reveal} those edges at the end of round one.)
	
	The same reasoning iterates: in round two, the “effective” shortest length becomes $\sqrt{L}\gamma$, and so on. 
	After $i$ rounds the algorithm has essentially learned only the edges belonging to cycles of length at most $\sqrt{L}\,\gamma^{i}$.
	
	To eliminate all cycles (and hence recover a spanning forest), the algorithm must “reach” cycles of length $\Theta(L)$, i.e., require $\sqrt{L}\,\gamma^i \approx L$. This means we need $i \approx \log_{\gamma}(\sqrt{L}) = \Omega(\log L) = \Omega(\log m)$ rounds.
		Hence any algorithm restricted to $\mathrm{poly}(m)$ queries per round needs $\Omega(\log m)$ rounds on this instance family.
	
	For the formal proof, one must also account for information leaked by \emph{independent} queries (which certify that many edge sets contain no cycles, including longer ones). We bound this information carefully, showing it does not accelerate progress beyond the shortest-cycle layer in each round; chaining these bounds yields the claimed $\Omega(\log m)$ lower bound.
	
	\subsection{Organization}
	
	In \cref{sec:upperBoundRandom}, we present the formal analysis of our cycle overlap counting bound, our sampling procedure for enumerating cycles, and our procedure for removing cycles, culminating in a $\mathrm{poly}(m)$ query, $O(\log(m))$ round randomized algorithm for finding spanning forests. In \cref{sec:derand}, we show how to derandomize the cycle enumeration algorithm, and in \cref{sec:lowerbound}, we present the formal analysis for our lower bound. In \cref{sec:otherMatroids}, we generalize the basis finding algorithm to broader classes of matroids. 
	
	\section{Randomized Algorithm for Graphic Matroids}\label{sec:upperBoundRandom}
	
	We start by covering some auxiliary lemmas that will be helpful in the analysis of our algorithm.
	
	\subsection{Cycle Counting Bounds}
	
	To start, recall the statement from the work of \cite{Sub95}\footnote{This work only shows the bound for a constant multiple of the minimum cycle length. A simple adaptation holds for arbitrary values times the minimum cycle length.}:
	
	\begin{claim}[\cite{Sub95}]
		Let $G$ be a graph on $n$ vertices, and suppose that $G$ has no cycles of length $\leq \ell$. Then, $G$ has at most $n^4$ cycles of length $\leq 2 \ell$.
	\end{claim}
	
	In fact, we will need a stronger version of this statement (which has appeared in the literature, see \cite{GV19} for instance).
	
	\begin{claim}\label{clm:cycleBound}
		Let $G$ be a graph with $\leq m$ edges, and minimum cycle length $\ell$. Then, for any $\alpha \in \Z^+$, the number of cycles of length $\leq \alpha \cdot \ell$ is at most $(2m)^{2\alpha}$.
	\end{claim}
	
	Next, we use this counting bound to bound the \emph{overlap} between cycles.
	
	\subsection{Cycle Overlap Properties}
	
	To better understand properties of cycle overlaps, we need the following claims and lemmas. To start, we have the following:
	
	\begin{claim}[Symmetric Difference of Cycles]\label{clm:disjointUnionCycles}
		Let $C, C'$ be two distinct cycles in a graph. Then the symmetric difference of $C, C'$ (denoted by $C \oplus C'$) contains a cycle.
	\end{claim}
	
	\begin{proof}
		Consider the set of edges given by $C \oplus C'$, and let us consider the degree of any vertex $v$. We claim that $v$ will always have degree $0, 2$ or $4$ in $C \oplus C'$, and thus there must be a cycle in the graph $C \oplus C'$ (as it is a non-empty graph with all degrees being even). 
		
		To see why, if $v$ has degree $0$ in both $C, C'$, then $v$ also has degree $0$ in $C \oplus C'$. If $v$ has degree $2$ in $C$, but degree $0$ in $C'$ (or vice versa), then $v$ will have degree $2$ in $C \oplus C'$, as no edges incident to $v$ are removed. Finally, if $v$ has degree $2$ in both $C$ and $C'$, then we must look at how many edges are in common. If $v$ shares $0$ edges between $C, C'$ then its degree in the symmetric difference is $4$, if $v$ shares $1$ edge between $C, C'$, its degree is $2$, and if it shares both edges between $C, C'$, then its degree is $0$ in the symmetric difference. This concludes the proof. 
	\end{proof}
	
	Next, we establish the following lemma, which lower bounds the overlap size between any cycles: 
	
	\begin{lemma}\label{lem:cycleOverlapGeneral}
		Let $E$ be a set of edges such that there is no cycle of length $\leq \ell$. Let $C$ be a cycle of length $k \leq 1.01 \ell$ in $E$ and let $C'$ be a cycle of length $k'$ in $E$. Then, $|C' \setminus C| \geq \frac{k'}{4}$.
	\end{lemma}
	
	\begin{proof}
		Note that if $k'$ is sufficiently large relative to $\ell$, the above lemma is trivial. For instance, if $k' \ge 1.5\ell$, then $|C' \setminus C| \ge k' - 1.01\ell \ge k' - (3/4)k' = \frac{k'}{4}$ (using $1.01\ell \le (3/4)k'$ for $k' \ge 1.5\ell$). The interesting case is when $k' \le 1.5\ell$. Here we use Claim~\ref{clm:disjointUnionCycles}: since $C \oplus C'$ must contain a cycle and $\ell$ is the shortest cycle length in $G$, we have $|C \oplus C'| \ge \ell$. We can rewrite 
		\[|C \oplus C'| = |C| + |C'| - 2|C \cap C'| = |C| + |C'| - 2(|C'| - |C' \setminus C|).\] 
		Plugging in $|C \oplus C'| \ge \ell$ and using $|C| \le 1.01\ell$ and $|C'| \ge \ell$, we obtain 
		\[ |C| + |C'| - 2(|C'| - |C' \setminus C|) \ge \ell, \] 
		which simplifies to 
		\[ 2|C' \setminus C| \ge \ell + |C'| - |C| \ge 0.99\ell. \] 
		Thus $|C' \setminus C| \ge 0.495\ell$, which in this case is at least $k'/4$. 
	\end{proof}
	
	With these lemmas, we are now ready to start presenting our algorithms. We begin by studying algorithms for recovering unique cycles in graphs using only an independence oracle.
	
	\subsection{Detecting a Single Cycle}
	
	As mentioned in the introduction, our algorithm proceeds by removing cycles in an iterative manner, gradually eliminating cycles of increasing lengths. Suppose in some iteration of the algorithm we have the promise that there are no cycles of length $\leq \ell$ in the graph. Then our goal for the iteration is to (1) eliminate all cycles of length $\leq 1.01 \ell$ and (2) to do this \emph{without} altering the connectivity of the graph. We will accomplish this task by repeatedly sampling the edges in the graph to create the following {\em good event}: that there is a unique cycle that survives among the sampled edges and moreover, its length is $\leq 1.01 \ell$.  
	
	Conditioned on this good even, we must then identify exactly the edges that are participating in this unique surviving cycle, and then repeat this process many times until we have enumerated all cycles of length $\leq 1.01 \ell$ in the graph. Note that because of \cref{clm:cycleBound}, as a sanity check we can see that the number of potential cycles we must recover is bounded by some polynomial in $m$ (although there is no guarantee that these cycles are easy to find). 
	
	As a first step towards identifying these cycles, we present a simple algorithm for detecting whether or not there is a single cycle in a graph (and if there is a single cycle, the algorithm returns exactly the edges in the cycle):
	
	\begin{algorithm}[H]
		\caption{DetectSingleCycle($E'$)}\label{alg:detectSingleCycle}
		Initialize the set of critical edges $S = \emptyset$.\\
		\If{\text{Query }$\mathrm{Ind}(E') = 1$}{
			\Return{$\perp$, No cycles.}
		}
		\For{$e \in E'$}{
			\If{\text{Query }$\mathrm{Ind}(E' \setminus \{e\}) = 1$}{
				$S \leftarrow S \cup \{ e \}$.
			}
		}
		\If{$S = \emptyset$}{
			\Return{$\perp$, $\geq 2$ cycles.}
		}
		\Return{$S$.}
	\end{algorithm}
	
	We summarize the performance of the algorithm in several claims:
	
	\begin{claim}\label{clm:detectSingleCycle1}
		Let $E'$ be a set of edges which has no cycles, then \cref{alg:detectSingleCycle} correctly returns that there are no cycles.
	\end{claim}
	
	\begin{proof}
		The independence query to $E'$ will be $1$ if and only if there are no cycles.
	\end{proof}
	
	\begin{claim}\label{clm:detectSingleCycle2}
		Suppose $E'$ has $\geq 2$ cycles, then \cref{alg:detectSingleCycle} returns that there are $\geq 2$ cycles.
	\end{claim}
	
	\begin{proof}
		To prove this, we use an auxiliary claim: namely that if a graph $G$ has at least $2$ distinct cycles, then there is no single edge whose removal kills both cycles. To see why, let us denote two cycles by $C_1, C_2$. If an edge $e$ is in both $C_1$ and $C_2$, then by \cref{clm:disjointUnionCycles}, $C_1 \oplus C_2$ contains a cycle, \emph{and} $e \notin C_1 \oplus C_2$. Thus, there must be some cycle in the graph which does not include the edge $e$, so $e$'s removal does not disconnect the graph. Otherwise, if $e$ is not in both of $C_1, C_2$, one of $C_1, C_2$ will also remain intact after $e$'s removal. This yields the claim.
	\end{proof}
	
	\begin{claim}\label{clm:detectSingleCycle3}
		Suppose $E'$ has exactly $1$ cycle, then \cref{alg:detectSingleCycle} returns exactly the constituent edges of this cycle. 
	\end{claim}
	
	\begin{proof}
		Let the cycle be denoted by $C$. Observe that removing any edge $e \in C$ will disconnect the cycle, and hence the independence queries will now return $1$. Likewise, deleting any edge $e \notin C$ will not disconnect the cycle, and the cycle will still be present, so the independence queries will return $0$. 
	\end{proof}
	
	Finally, we also observe that the above algorithm can be implemented in $1$ round of adaptivity:
	
	\begin{claim}\label{clm:detectSingleCycle4}
		\cref{alg:detectSingleCycle} can be implemented in $1$ round of adaptivity and makes $|E'| + 1$ queries to $\mathrm{Ind}$.
	\end{claim}
	
	\begin{proof}
		Notice that the algorithm queries $\mathrm{Ind}(E')$, and $\mathrm{Ind}(E' \setminus \{e\}): \forall e \in E'$. All these queries are made without reference to the results from previous queries.
	\end{proof}
	
	Thus, we get the following lemma to summarize \cref{alg:detectSingleCycle}:
	
	\begin{lemma}\label{lem:detectSingleCycle}
		For a set of edges $E'$, \cref{alg:detectSingleCycle} makes $|E'| +1$ queries to $\mathrm{Ind}$ in only a single round, and returns $\perp$ if $E'$ has no cycles or $\geq 2$ cycles, and otherwise returns a set $S \subseteq E'$ which is exactly the edges involved in the single cycle in $E'$.
	\end{lemma}
	
	\begin{proof}
		This follows from \cref{clm:detectSingleCycle1}, \cref{clm:detectSingleCycle2}, \cref{clm:detectSingleCycle3}, and \cref{clm:detectSingleCycle4}.
	\end{proof}
	
	Thus, we have a simple algorithm for identifying when there is a single cycle in a graph. In the coming sections, we will show how to use this procedure to identify and remove all of the short cycles in a given graph.
	
	\subsection{Sampling}
	
	Motivated by the previous subsection, our goal will now be to sub-sample the graph at a specific rate such that only \emph{one} cycle will survive the sampling process with high probability. When only one cycle survives, we can then identify this cycle exactly by \cref{alg:detectSingleCycle}. The sampling algorithm is provided below, which takes in a number of edges $m$, the edge set $E$, as well as a parameter $\ell$ corresponding to the minimum cycle length in $E$.
	
	\begin{algorithm}[H]
		\caption{RecoverCycleSuperset$(E, m, \ell)$}\label{alg:recover}
		Let $p$ be such that $p^{\ell} = \frac{1}{m^{100}}$. \\
		\For{$i \in [m^{102}]$, in parallel }{
			Let $E^{(i)}$ be the result of independently keeping each $e \in E$ with probability $p$.
		}
		\Return{$\{E^{(i)}: i \in [m^{102}]\}$}
	\end{algorithm}

	To understand the sampling procedure, we focus on a single cycle $C$ in $E$ of length $k$, for $k \in (\ell, 1.01 \ell]$, and start by showing the following:
	
	\begin{claim}\label{clm:uniqueCycleProb}
		Let $E$ be a set of $m$ edges such that there is no cycle of length $\leq \ell$ and let $C$ be a cycle in $E$ of length $k$, for $k \in (\ell, 1.01 \ell]$. Let $E'$ be the result of independently keeping each edge in $E$ with probability $p$, where $p^{\ell} = \frac{1}{m^{100}}$. Then, $C$ is the unique cycle in $E'$ with probability $\frac{1}{2\cdot m^{101}}$.
	\end{claim}
	
	\begin{proof}
		First, we calculate the probability that $C$ survives the sampling procedure. This is straightforward, as $C$ has $\leq 1.01 \ell$ edges. So, after sampling at rate $p$, the probability all $\leq 1.01 \ell$ edges survive is:
		\[
		\Pr[C \text{ survives sampling}] \geq p^{1.01 \ell} = \left ( p^{\ell} \right )^{1.01} = \left ( \frac{1}{m^{100}} \right )^{1.01} = \frac{1}{m^{101}}.
		\]
		
		Thus, it remains only to bound the probability that \emph{some other cycle} $C'$ also survives the sampling at rate $p$, conditioned on $C$ surviving the sampling. 
		
		Indeed, to bound this, let us suppose that $C'$ is of length $k'$. By \cref{lem:cycleOverlapGeneral}, we know that $|C' \setminus C| \geq \frac{k'}{4}$. Thus,
		\[
		\Pr[C' \text{ survives sampling}  | C \text{ survives sampling}] \leq p^{k'/4},
		\]
		as there will be at least $k'/4$ edges in $C'$ which are not in $C$ (and thus these edges surviving the sampling process is independent of $C$ surviving). 
		
		Next, it remains to take a union bound over all possible cycles $C'$. For this, let $\alpha$ be the power of $2$ such that $k' \in [\alpha \ell, 2\alpha \ell]$, and then we use \cref{clm:cycleBound}. Specifically, for cycles $C'$ of length $[\alpha \ell, 2\alpha \ell]$, we know that there are at most $(2m)^{2\alpha}$ such cycles. So, for a fixed $\alpha$, this means we get the following bound:
		\[
		\Pr[\exists C'\neq C \text{ of length }[\alpha \ell, 2\alpha \ell] \text{ that survives sampling} | C \text{ survives sampling}] 
		\]
		\[
		\leq \sum_{C'\neq C: \text{ cycle of length }[\alpha \ell, 2\alpha \ell]} \Pr[C' \text{ survives sampling}  | C \text{ survives sampling}]
		\]
		\[
		\leq \sum_{C': \text{ cycle of length }[\alpha \ell, 2\alpha \ell]} p^{\alpha \ell/4} \leq p^{\alpha \ell/4} \cdot (2m)^{2 \alpha} = \left ( \frac{1}{m^{100}} \right )^{\frac{\alpha}{4}} \cdot (2m)^{2\alpha} \leq \left (\frac{2}{m^{23}} \right)^{\alpha}.
		\]
		
		To conclude, we can then take a union bound over $\alpha \in \{1, 2, 4, 8, \dots m \}$. Thus, we see that:
		\[
		\Pr[\exists C'\neq C\text{ that survives sampling} | C \text{ survives sampling}] 
		\]
		\[
		\leq \sum_{\alpha \in \{1, 2, 4, 8, \dots m \}} \Pr[\exists C'\neq C \text{ of length }[\alpha \ell,2\alpha \ell] \text{ that survives sampling} | C \text{ survives sampling}] 
		\]
		\[
		\leq \sum_{\alpha \in \{1, 2, 4, 8, \dots m \}} \left (\frac{2}{m^{23}} \right)^{\alpha} \leq \sum_{\alpha = 1}^{\infty} \left (\frac{2}{m^{23}} \right)^{\alpha} 
		\]
		\[
		\leq \frac{4}{m^{23}}
		\]
		where the final inequality follows because the expression is a geometric series with ratio $< 1/2$.
		
		Finally, recall that our goal was to show that $C$ is the unique cycle which survives sampling with non-negligible probability. For this, observe that 
		\[
		\Pr[C \text{ survives sampling}] = \Pr[C \text{ uniquely survives sampling}] + \Pr[\exists C'\neq C: C \wedge C' \text{ survive sampling}].
		\]
		Now, this second term we can bound by our above work. I.e.,
		\[
		\Pr[\exists C'\neq C: C \wedge C' \text{ survive sampling}]
		\]
		\[
		= \Pr[\exists C'\neq C\text{ that survives sampling} | C \text{ survives sampling}]\cdot\Pr[C \text{ survives sampling}]
		\]
		\[
		\leq \frac{4}{m^{23}} \cdot \Pr[C \text{ survives sampling}].
		\]
		Thus, we see that 
		\[
		\Pr[C \text{ uniquely survives sampling}] \geq \Pr[C \text{ survives sampling}] \cdot \left ( 1 - \frac{4}{m^{23}}\right ) \geq \frac{1}{2\cdot m^{101}},
		\]
		as we desire.
	\end{proof}
	
	Using the above claim, we now show that (with high probability) every cycle $C$ of length $(\ell, 1.01 \ell]$ is the unique surviving cycle for some $E^{(i)}$ produced by \cref{alg:recover}.
	
	\begin{lemma}\label{lem:uniqueCycleSurvive}
		Let $E$ be a set of $m$ edges such that there is no cycle of length $\leq \ell$. Then, with probability $1 - 2^{-\Omega(m)}$, for every cycle $C$ in $E$ of length $(\ell, 1.01 \ell]$, there is an index $i \in [m^{102}]$ such that $C$ is the unique cycle in $E^{(i)}$.
	\end{lemma}
	
	\begin{proof}
		Fix any cycle $C$ in $E$ of length $(\ell, 1.01 \ell]$. By \cref{clm:uniqueCycleProb}, we know that over the randomness of the sampling procedure, $C$ will be the unique cycle present in $E^{(i)}$ with probability $\frac{1}{2m^{101}}$. Thus, by repeating this procedure $m^{102}$ times, we know that there is at least one index $i$ for which $C$ is the unique cycle with probability $1 - 2^{-\Omega(m)}$.
		
		Now, because there are at most $(2m)^5$ cycles of length $(\ell, 1.01\ell]$ in a graph with no cycles of length $\leq \ell$, we can take a union bound over all these cycles. This means that with probability $1 - 2^{-\Omega(m)}$, for every cycle $C$ in $E$ of length $(\ell, 1.01 \ell]$, there is an index $i \in [m^{102}]$ such that $C$ is the unique cycle in $E^{(i)}$.
	\end{proof}

	\subsection{Final Cycle Recovery Algorithm}
	
	Now, it remains only to piece together these algorithms. We present this below as an algorithm:
	
	\begin{algorithm}[H]
		\caption{TotalCycleRecovery$(E, m, \ell)$}\label{alg:final}
		Let $p$ be such that $p^{\ell} = \frac{1}{m^{100}}$. \\
		$\mathrm{Cycles}= \{\}$.\\
		\For{$i \in [m^{102}]$ in parallel}{
			Let $E^{(i)}$ be the result of independently keeping each $e \in E$ with probability $p$. \\
			Let $S = \mathrm{DetectSingleCycle}(E^{(i)})$. \\
			\If{$S \neq \perp$}{
				$\mathrm{Cycles} \leftarrow \mathrm{Cycles} \cup \{S\}$.
			}
		}
		\Return{$\mathrm{Cycles}$.}
	\end{algorithm}
	
	We can combine our results from the previous subsections to understand what \cref{alg:final} achieves:
	
	\begin{lemma}\label{lem:recoverShortCycles}
		Let $E$ be a set of $m$ edges such that there is no cycle of length $\leq \ell$. Then, with probability $1 - 2^{-\Omega(m)}$, the output of \cref{alg:final} is a set of cycles which contains every cycle $C$ in $E$ of length $(\ell, 1.01 \ell]$.
	\end{lemma}
	
	\begin{proof}
		First, recall that by \cref{lem:detectSingleCycle}, the algorithm DetectSingleCycle does not return $\perp$ if and only if there is a single cycle in the input graph. By this same lemma, when the output is not $\perp$, the algorithm recovers exactly the constituent edges of the cycle. Hence the output of \cref{alg:final} is necessarily a set of cycles (as the cycles in the subsampled graphs will also be cycles in the original graph). It remains only to show that \emph{every} cycle of length $(\ell, 1.01 \ell]$ is included in the output. This follows exactly from \cref{lem:uniqueCycleSurvive}. Indeed, every cycle of length $(\ell, 1.01 \ell]$ will be the unique cycle surviving the sampling procedure with probability $1 - 2^{-\Omega(m)}$, and thus will be included in the output as well. This yields the lemma at hand. 
	\end{proof}
	
	Likewise, we can observe that \cref{alg:final} is implementable in parallel without any adaptivity:
	
	\begin{claim}
		On input $E$, with $m$ edges, and parameter $\ell$, \cref{alg:final} is implementable with $\mathrm{poly}(m)$ queries to the independence oracle in a single round.
	\end{claim}
	
	\begin{proof}
		By \cref{lem:detectSingleCycle}, each invocation of DetectSingleCycle requires $O(m)$ queries to the independence oracle. Each of the sub-sampled graphs is checked in parallel, and thus the total number of queries is $O(m \cdot m^{102}) = \mathrm{poly}(m)$, without any adaptivity. 
	\end{proof}
	
	\subsection{Cycle Removal}
	
	After recovering all of the cycles of length $(\ell, 1.01\ell]$, we must find a way to delete these cycles without altering the connectivity of the graph. To do this, we take advantage of a basic operation from the work of \cite{KUW85}. Although this statement is included in \cite{KUW85}, there was no proof provided there, so we re-prove the result here.
	
	\begin{lemma}\label{lem:deleteCycles}
		Let $E$ be a set of edges with some fixed ordering of the edges $e_1, \dots e_m$, and let $\mathrm{Cycles}$ be an arbitrary subset of the cycles in $E$. For each cycle $C \in \mathrm{Cycles}$, let $C = (e_{i_{C, 1}}, \dots e_{i_{C, |C|}})$ denote the ordered set of the edges that are in the cycle $C$. Let $E'$ be the result of simultaneously deleting from $E$ the edge with the \emph{largest} index from every cycle in $\mathrm{Cycles}$. Then,
		\begin{enumerate}
			\item Every cycle in $\mathrm{Cycles}$ has at least one edge removed.
			\item The connected components of $E'$ are the same as the connected components of $E$. 
		\end{enumerate}
	\end{lemma}
	
	\begin{proof}
		The first item is essentially trivial. Every cycle in $\mathrm{Cycles}$ has some constituent edge deleted.
		
		The second item is less immediate. To see why the connectivity of $E$ and $E'$ is the same, let us consider adding the edges in $E \setminus E'$ back in to $E'$, in increasing order (i.e., starting with the edge which has the smallest label). Let us denote these edges by $e_1^*, \dots e_{g}^{*}$. We claim that the connectivity of $E' \cup \{e_1^*, \dots e_{i}^* \}$ is the same as the connectivity of $E' \cup \{e_1^*, \dots e_{i+1}^* \}$. For the base case, we show that the connected components of $E'$ are the same as the connected components of $E' \cup \{e^*_1 \}$. This is because $e^*_1$ was the smallest labeled edge which was removed. Because it was removed though, this means that there is some cycle $C_1$ for which $e^*_1$ has the largest label, and hence all other edges in $C_1 \setminus \{ e^*_1\}$ are still present in $E'$. But, adding $e^*_1$ then does not alter the connected components, as the vertices connected by $e^*_1$ are already connected by the edges in $C_1 \setminus \{ e^*_1\}$. 
		
		Now, we show the general case. I.e., that the connectivity of $E' \cup \{e_1^*, \dots e_{i}^* \}$ is the same as the connectivity of $E' \cup \{e_1^*, \dots e_{i+1}^* \}$. For this, observe that because the $e_j^*$'s are ordered in terms of their labels, $E' \cup \{e_1^*, \dots e_{i}^* \}$ contains every edge whose label is smaller than $e_{i+1}^*$. In particular, because $e_{i+1}^*$ was removed, this means that $e_{i+1}^*$ was the largest labeled edge in some cycle $C_{i+1}$. This means that every edge in $C_{i+1} \setminus \{ e_{i+1}^*\}$ is still in $E' \cup \{e_1^*, \dots e_{i}^* \}$. But, adding $e^*_{i+1}$ then does not alter the connected components, as the vertices connected by $e^*_{i+1}$ are already connected by the edges in $C_{i+1} \setminus \{ e_{i+1}^*\}$. 
		
		To conclude, we simply observe that by induction, this means that the connected components of $E'$ are the same as the connected components of  $E' \cup \{e_1^*, \dots \} = E$.
	\end{proof}
	
	\subsection{Spanning Forest Computation}
	
	With all of our building blocks now established, we present our final algorithm for finding a spanning forest below. Note that because we make progress on the order of $\ell \rightarrow 1.01\ell$ in each round, we need to initialize our algorithm on a graph with no cycles of length $\leq 100$. To do this, we enumerate all cycles of length $\leq 100$ explicitly in the first round:
	
	\begin{algorithm}[H]
		\caption{FindSpanningForest$(E, m)$}\label{alg:findSpanningForest}
		Let $e_1, \dots e_m$ be a fixed ordering of the edges of $E$. \\
		$\ell = 100$. \\
		$\mathrm{Cycles} = \emptyset$. \\
		\For{$E' \subseteq E: |E'| \leq \ell$ in parallel}{
			$S = \mathrm{DetectSingleCycle}(E')$.\\
			\If{$S\neq \perp$}{
				$\mathrm{Cycles} \leftarrow \mathrm{Cycles} \cup \{S\}$
			}
		}
		\For{$C \in \mathrm{Cycles}$ in parallel}{
			Let $e^*$ be the edge in $C$ with the largest index with respect to the ordering. \\
			$E \leftarrow E \setminus \{ e^*\}$.
		}
		\While{$|E| \geq \ell$}{
			$\mathrm{Cycles} = \mathrm{TotalCycleRecovery}(E, |E|, \ell)$. \\
			\For{$C \in \mathrm{Cycles}$ in parallel}{
				Let $e^*$ be the edge in $C$ with the largest index with respect to the ordering. \\
				$E \leftarrow E \setminus \{ e^*\}$.
			}
			
			$\ell \leftarrow \lfloor 1.01 \cdot \ell \rfloor$.
		}
		\Return{$E$.}
	\end{algorithm}
	
	We now present a sequence of claims analyzing the above algorithm.
	
	\begin{claim}\label{clm:preprocess}
		After invoking \cref{alg:findSpanningForest} on a set of edges $E$ with $m$ edges, after Line 13, the connected components of $E$ have not changed and there are no cycles left of length $\leq 100$.
	\end{claim}
	
	\begin{proof}
		Observe that the set of cycles in line 10 is indeed a set of cycles as per \cref{lem:detectSingleCycle}. In particular, every cycle of length $\leq 100$ will be recovered, as the algorithm queries all subsets of $\leq 100$ edges, and so there will be some query for each cycle. The cycle deletion procedure is exactly that of \cref{lem:deleteCycles}, and thus all cycles of length $\leq 100$ are removed, without altering the connected components of $E$. 
	\end{proof}
	
	\begin{claim}\label{clm:singleIteration}
		Let $E$ be a set of $m$ edges with no cycles of length $\leq \ell$. For each iteration of the while loop (Line 14) in \cref{alg:findSpanningForest}, the algorithm removes all cycles of length $\leq 1.01 \ell$ without altering the connected components of $E$, with probability $1 - 2^{-\Omega(m)}$.
	\end{claim}
	
	\begin{proof}
		This set of cycles recovered includes all cycles of length $\leq 1.01 \ell$ with probability $1 - 2^{-\Omega(m)}$ by \cref{lem:recoverShortCycles}. These cycles are then deleted without altering the connected components as per \cref{lem:deleteCycles}.
	\end{proof}
	
	Next, we also bound the number of iterations of the while loop:
	
	\begin{claim}\label{clm:numberRounds}
		After invoking \cref{alg:findSpanningForest} on a set of edges $E$ with $m$ edges, the while loop in line 14 runs for at most $O(\log(m))$ iterations, with probability $1 - 2^{-\Omega(\sqrt{m})}$, and returns a set of edges with no cycles of length $\leq m+1$.
	\end{claim}
	
	\begin{proof}
		First, recall that any connected component on $k$ vertices has $\leq k^2$ edges. Thus, for a set $E$ of $m$ edges, any spanning forest must have $\geq \sqrt{m}$ edges. Thus, as long as the set of connected components defined by $E$ has not changed under deleting edges, the number of remaining edges must be $\Omega(\sqrt{m})$.
		
		Next, observe that the set of edges $E$ in the input to line 14 has no cycles of length $\leq 100$ as per \cref{clm:preprocess} (but still has $\Omega(\sqrt{m})$ edges). Now, by \cref{clm:singleIteration}, with probability $1 - 2^{- \Omega(\sqrt{m})}$, all cycles of length $\leq 100 \cdot 1.01$ are removed. 
		
		In general, observe that for an integer $k \geq 100$, it must be the case that 
		\[
		\lfloor k \cdot 1.01 \rfloor \geq k \cdot 1.005.
		\]
		This is because 
		\[
		\lfloor k \cdot 1.01 \rfloor = k + \lfloor k \cdot 0.01 \rfloor,
		\]
		and for $k \geq 100$, we can write $k = \alpha \cdot 100 + k \mod 100$ for $\alpha \geq 1$. This means
		\[
		\lfloor k \cdot 0.01 \rfloor = \alpha,
		\]
		while 
		\[
		0.005 k = \alpha /2 + (k \mod 100) \cdot 0.005 < \alpha/2 + 1/2. 
		\]
		Thus,
		\[
		\lfloor k \cdot 1.01 \rfloor = k + \lfloor k \cdot 0.01 \rfloor = k + \alpha \geq k + \alpha/2 + 1/2 > k + 0.005 k = 1.005k.
		\]
		Thus, if we let $\ell$ denote the minimum cycle length for an iteration of the while loop in line 14, after this iteration, the new minimum cycle length will be $\geq 1.005 \ell$. After $O(\log(m))$ iterations, the minimum cycle length will be $\geq m+1$, and thus there will be no cycles remaining. In particular, this also means that the number of edges remaining (denoted by $|E|$) is less than the minimum cycle length, and so the stopping condition of the while loop is met. 
		
		Note that the probability bound holds because in each iteration there are $\Omega(\sqrt{m})$ edges remaining (where $m$ denotes the initial number of edges), and each iteration increases the cycle length by a factor of $\geq 1.005$ with probability $1 - 2^{-\Omega(\sqrt{m})}$. Thus, we can afford to take a union bound over each iteration failing. This yields the claim.
	\end{proof}
	
	We conclude with our primary theorem:
	
	\begin{theorem}
		\cref{alg:findSpanningForest} outputs a spanning forest of a graphic matroid on $m$ edges with probability $1 - 2^{-\Omega(\sqrt{m})}$ using $\mathrm{poly}(m)$ queries to an independence oracle per round and $O(\log(m))$ rounds of adaptivity.
	\end{theorem}
	
	\begin{proof}
		To see the bound on the number of queries, and rounds of adaptivity, recall that each invocation of $\mathrm{DetectSingleCycle}$ requires $O(m)$ queries to the oracle by \cref{lem:detectSingleCycle}. In each round, there are $\mathrm{poly}(m)$ invocations made to \cref{lem:detectSingleCycle} (in parallel), and thus the total number of queries per round is $\mathrm{poly}(m)$. Likewise, by \cref{clm:numberRounds}, the number of rounds required is $O(\log(m))$, and returns a spanning forest of $E$ with probability $1 - 2^{-\Omega(\sqrt{m})}$. This yields the theorem. 
	\end{proof}
	
	\section{Derandomization for Graphic Matroids}\label{sec:derand}
	
	In this section, we show that the algorithm from the previous section can be derandomized. To start, let us recall the statement of \cref{clm:uniqueCycleProb}:
	
	\begin{claim}[Restatement of \cref{clm:uniqueCycleProb}]
		Let $E$ be a set of $m$ edges such that there is no cycle of length $\leq \ell$ and let $C$ be a cycle in $E$ of length $k$, for $k \in (\ell, 1.01 \ell]$. Let $E'$ be the result of independently keeping each edge in $E$ with probability $p$, where $p^{\ell} = \frac{1}{m^{100}}$. Then, $C$ is the unique cycle in $E'$ with probability $\frac{2}{m^{101}}$.
	\end{claim}
	
	Our derandomization relies on the following observation: if we repeatedly randomly sample sets in accordance with \cref{clm:uniqueCycleProb}, then we can push the probability of not uniquely recovering a cycle to be exponentially small (for a large polynomial in the exponent). Then, we can simply observe that while the number of graphic matroids is exponential, the failure probability can be made so small that it survives a union bound over all graphic matroids. 
	
	To start, we establish the following claim:
	\begin{claim}
		Let $G$ be a graphic matroid over $m$ edges such that there is no cycle of length $\leq \ell$ and let $C$ be any cycle in $E$ of length $k$, for $k \in (\ell, 1.01 \ell]$. Let $B_1, \dots B_{m^{200}}$ be random sets, where each $B_i$ is the result of (independently) sampling $[m]$ at rate $p$, for $p^{\ell} = \frac{1}{m^{100}}$. Then, there exists an $i \in [m^{200}]$ such that $C$ is the unique surviving cycle in $B_i$ with probability $\geq 1 - 2^{-m^{98}}$.
	\end{claim}
	
	\begin{proof}
		Observe that by \cref{clm:uniqueCycleProb}, the cycle $C$ is the unique surviving cycle in the set $B_i$ with probability $\geq \frac{1}{2m^{101}}$. In particular, we can see that if we repeat this $m^{200}$ times, we get:
		\[
		\Pr[\exists i: C \text{ unique surviving in }B_i] \geq 1 - \left(1 - \frac{1}{2m^{101}}\right)^{m^{200}} \geq 1 - 2^{-m^{98}}.
		\]
		This yields the claim. 
	\end{proof}
	
	Now, we can take a simple union bound over all cycles:
	
	\begin{claim}\label{clm:fixedMatroidRandomSets}
		Let $G$ be a graphic matroid over $m$ edges such that there is no cycle of length $\leq \ell$. Let $B_1, \dots B_{m^{200}}$ be random sets, where each $B_i$ is the result of (independently) sampling $[m]$ at rate $p$, for $p^{\ell} = \frac{1}{m^{100}}$. Then, for every cycle $C$ of length $k$, for $k \in (\ell, 1.01 \ell]$, there exists an $i \in [m^{200}]$ such that $C$ is the unique surviving cycle in $B_i$ with probability $\geq 1 - (2m)^4 \cdot 2^{-m^{98}} \geq 1 - 2^{-m^{97}}$.
	\end{claim}
	
	\begin{proof}
		Note that if a graph has $m$ edges, it must be on $\leq 2m$ vertices. It follows then that there are $\leq (2m)^4$ cycles of length $\leq 1.01 \ell$. For each of these cycles, there exists a $B_i$ which isolates the cycle with probability $1- 2^{-m^{98}}$. Taking the union bound over these cycles then yields the claim. 
	\end{proof}
	
	\begin{claim}
		Let $m, \ell$ be integers. There exists a universal set of queries $B_1, \dots B_{m^{200}}$, where each $B_i$ is a subset of $[m]$ such that \emph{for any} graphic matroid $G$ on $m$ elements with minimum cycle length $\ell$, and any cycle $C$ in $G$ of length $[\ell, 1.01 \ell]$, there is a set $B_i$ for which $C$ is the unique surviving cycle in $B_i$.
	\end{claim}
	
	\begin{proof}
		By \cref{clm:fixedMatroidRandomSets}, we know that for a fixed graphic matroid $G$ and every cycle $C$ in $G$ of length $[\ell, 1.01\ell]$, a random selection of $B_1, \dots B_{m^{200}}$ will contain a $B_i$ for which $C$ is the unique surviving cycle with probability $\geq 1 - 2^{-m^{97}}$. When this condition holds, we say that $B_1, \dots B_{m^{200}}$ is \emph{good} for $G$.
		
		In particular, we can now take a union bound over all possible graphic matroids. Observe that if there are $m$ elements in the graphic matroid, the matroid must be defined on $\leq 2m$ vertices. The number of possible graphs with $m$ labelled edges on $2m$ vertices is at most
		\[
		\binom{2m}{2}^{m}.
		\]
		Note that this quantity above includes all possible permutations because we are using labelled edges. Thus, the total number of possible graphic matroids on $m$ elements is 
		\[
		\leq \binom{2m}{2}^{m} \leq 2^{O(m^2)}.
		\]
		
		We can then take a union bound over all possible graphic matroids. We obtain that a random choice of $B_1, \dots B_{m^{200}}$ will be \emph{good} for all possible graphic matroids on $m$ elements (with minimum cycle length $\ell$) with probability $\geq 1 - 2^{O(m^2)} \cdot 2^{-m^{97}} \geq 1 - 2^{-m^{95}}$. To conclude then, this implies that there must \emph{exist} a choice of $B_1, \dots B_{m^{200}}$ which is good for every graphic matroid on $m$ elements (with minimum cycle length $\ell$).
	\end{proof}
	
	With this, we now conclude with our main theorem:
	
	\begin{theorem}
		There is a deterministic algorithm that outputs a spanning forest of a graphic matroid on $m$ edges using $O(\log(m))$ rounds of adaptivity and $\mathrm{poly}(m)$ queries to an independence oracle per round .
	\end{theorem}
	
	\begin{proof}
		For each $j \in [m]$, and every choice of $\ell \in [m]$, the algorithm is non-uniformly provided with the sets $B^{j,\ell}_1, \dots B^{j, \ell}_{m^{200}}$, where these sets are defined to be the sets such that every cycle of length $[\ell, 1.01\ell]$ in a graphic matroid on $j$ elements is the unique surviving cycle in some $B^{j,\ell}_i$.
		
		The remainder of the algorithm is as in \cref{alg:findSpanningForest}, with the primary difference being that the sub-routine of \cref{alg:final} uses the sets $B^{j,\ell}_1, \dots B^{j, \ell}_{m^{200}}$ instead of randomly chosen sets. As in \cref{alg:final}, having queries which uniquely isolate certain cycles allows the algorithm to find exactly the subset of edges that participate in the cycle, and so all cycles of length $[\ell, 1.01\ell]$ can be enumerated. Then, the analysis of the correctness of the algorithm follows \cref{clm:numberRounds}, as we iteratively remove cycles of length $[\ell,1.01\ell]$, and then $[1.01\ell,1.01^2 \ell]$ and so on. Observe that after $i$ rounds, the minimum cycle length is roughly $\ell = 1.01^{i}$, and the number of remaining edges $j$ is arbitrary, but $\leq m$ (the starting number of edges). Since the algorithm is provided with the sets $B^{j,\ell}$ for every choice of $j \leq m, \ell \leq m$, the algorithm will proceed and recover all cycles of length $[\ell,1.01\ell]$. This concludes the theorem. 
	\end{proof}
	
	\section{Lower Bound for Graphic Matroids}\label{sec:lowerbound}
	
	First, as in \cite{KUW85}, we present the idealized model for which we will create the lower bound. This is called the \emph{probabilistic parallel decision tree} of parallelism $q$ and oracle $f$. Such a tree consists of three types of nodes:
	
	\begin{enumerate}
		\item Randomization nodes, where an internal node splits into $b$ branches, each with probability $1 / b$.
		\item Oracle query nodes, where each node contains a set of queries $B_1, \dots B_q$, where each $B_i \subseteq E$. The branches from these nodes are then labeled by the possible oracle answers to these queries (denoted $f(B_1), \dots f(B_q))$.
		\item  Leaf nodes, which simply contain a subset of $E$ (with the goal that this subset is a spanning forest of the graph). 
	\end{enumerate}
	
	Recall that our goal is for the output node of such a probabilistic parallel decision tree to be a spanning forest of our graph with high probability. This is formalized by saying that a tree $H$ with the independence oracle $\mathrm{Ind}$ solves the spanning forest independence oracle problem if for any graphic matroid $G$ on $m$ elements, and any root to leaf path in $H$, if the leaf is labeled with set $B$, then $B$ is an independent set (i.e., no cycles), and has the same rank as the entire graph. In the upper bound, we showed that there is a probabilistic parallel decision tree whose expected depth is bounded by $O(\log(m))$ (before reaching a leaf node). We quantify this depth by setting $c(H, G)$ to be the expected depth of the computation reached by the tree $H$ when the input graph is $G$. The final quantity we will be interested in is $c(H, m) = \max_{G: |G| = m} c(H,G)$, i.e., the maximum expected depth over all possible input graphs. Our quantity of interest is then exactly $\min_H c(H, m)$, i.e., the tree which minimizes this maximum expected depth. We denote this quantity by $T_{\mathrm{prob}}^{\mathrm{ind}}(m,q)$.  We will show the following theorem:
	
	\begin{theorem}\label{thm:mainLB}
		Let $m$ be an integer denoting the number of edges, and let $q \leq m^c$ denote the number of independence queries in each round, for some constant $c > 0$. Then, $T_{\mathrm{prob}}^{\mathrm{ind}}(m,q) = \Omega(\log(m))$.
	\end{theorem}
	
	As a simplifying first step (and as observed in \cite{KUW85}), we can use Yao's minimax theorem to simplify our setting:
	
	\begin{claim}
		Let $T_1$ be the expected running time for a given probabilistic algorithm solving problem $P$, maximized over all possible inputs. Let $T_2$ be the average running time for a given input distribution, minimized over all possible deterministic algorithms that solve $P$. Then $T_1 \geq T_2$.
	\end{claim}
	
	Thus, instead of considering the \emph{randomized parallel decision tree} model discussed above, we can instead consider \emph{deterministic parallel decision trees}. Here, there are simply no randomization nodes. We can analogously define $T_{\mathrm{det}}^{\mathrm{ind}}(m,q)$ as the optimal deterministic computation time. To lower bound $T_{\mathrm{det}}^{\mathrm{ind}}(m,q)$, we must only present a distribution $D$ over graphs such that 
	\[
	T_{\mathrm{det}}^{\mathrm{ind}}(m, q, D) = \min_{H} \E[c(H, m)] = \Omega(\log(m))
	\]
	(conditioned on $q$ being polynomially bounded). Thus, we are now ready to present our distribution over graphs:
	
	\begin{definition}\label{def:instance}
		Let $q \leq m^{c/3}$, for some constant $c > 0$. The graphic matroid $G$ will depend on a parameter $\gamma = 2000 \cdot c^2$. The graph consists of $L = m^{1/2}/ \kappa$ (for $\kappa$ a large constant) disjoint cycles of length $\sqrt{L} \cdot \gamma$, $L$ disjoint cycles of length $\sqrt{L} \cdot \gamma^2$, and so on. Specifically, for $i \in [\log_{\gamma}(L)/2]$, the graph has $L$ disjoint cycles of length $\sqrt{L} \cdot \gamma^i$. Observe that the number of edges in the graph is $ O(L^2) \leq m$ (and can be padded with disjoint edges if desired). The randomness in the graph is taken over all possible labelings of the edges of the graph (i.e., all permutations on $[m]$). We denote this distribution over graphic matroids by $D$.
	\end{definition}
	
	Now, let us consider a fixed deterministic decision tree $H$ for computing a spanning forest of a graph. At each level of the tree, there are $q$ queries $B_1, \dots B_q$ that are performed, yielding a sequence of answers $\mathrm{Ind}(B_1), \dots \mathrm{Ind}(B_q)$. We let $O_1  = D$ denote the starting set of (all) possible permutations. More generally, we let $O_i$ denote the subset of permutations which agrees with the answers to queries in the first $i-1$ rounds.
	
	Next, as in \cite{KUW85}, we can observe that after each round of queries, we can without loss of generality reveal more information to the decision tree $H$. Thus, after the $i$th round of queries, we reveal to the decision tree $H$, the exact identity of all edges contained in cycles of length $\sqrt{L} \cdot \gamma^i$. We denote these sets of edges by $A_i$, and we denote by $Q_i$ the set of permutations which agree with the labels given to edges in $A_i$. Thus, as remarked in \cite{KUW85}, the probability of a given matroid (i.e., a given permutation over labels) at the start of the $i$th round is exactly the probability of the event $O_i \cap Q_i$ (under the uniform distribution $D$). 
	
	Now, we introduce another definition:
	
	\begin{definition}\label{def:local}
		We say that an oracle query $\mathrm{Ind}(B)$ in the $j$th step is \emph{local}: 
		\begin{enumerate}
			\item If $\mathrm{Ind}\left (B \cap \left ( \bigcup_{k \geq j} A_k \right )\right) = 0$, then $B$ has a cycle in $A_j$.
			\item Otherwise, if $\mathrm{Ind}\left (B \cap \left ( \bigcup_{k \geq j} A_k \right )\right) = 1$, then for $i > j$, 
			\[
			\left |B \cap \left ( \bigcup_{k \geq i} A_k \right ) \right | \leq \left ( 1 - \frac{100 c \log(L)}{\sqrt{L}\gamma^{i}}\right ) \cdot \left | \left ( \bigcup_{k \geq i} A_k \right ) \right |
			\]
		\end{enumerate}
	\end{definition}
	
	As in \cite{KUW85}, we will show that if the algorithm only performs local queries in the first $i-1$ rounds, then with high probability, the queries in the $i$th round will also be local. We formalize this below:
	
	\begin{claim}\label{clm:singleRoundLocal}
		For any $i \leq \log_{\gamma}(L)$, if all queries in the first $i-1$ rounds are local, then with probability $ 1- 3L^{-4c}$, all the queries in the $i$th round are also local. 
	\end{claim}
	
	\begin{proof}
		We let $E$ denote the event that at least one of the queries in the $i$th round is not local. Immediately, we can observe that 
		\[
		\Pr[E | O_i \cap Q_i] = \frac{\Pr[E \cap O_i | Q_i]}{\Pr[O_i | Q_i]} \leq \frac{\Pr[E| Q_i]}{\Pr[O_i | Q_i]}.
		\]
		
		Thus, our goal is to lower bound $\Pr[O_i | Q_i]$ and upper bound $\Pr[E| Q_i]$. We start by lower bounding $\Pr[O_i | Q_i]$. 
		
		Indeed, recall that $Q_i$ is simply the event that the edge labels exactly match the revealed edge labels in the cycles given by $A_{i-1}$. Importantly, once we condition on $Q_i$, the remaining distribution over the labels of the edges in $G \setminus {A_{i-1}}$ is uniformly random. Now, let us consider a query $B$ in the $j$th round, for $j < i$. Recall that every query is a local query by our assumption. Thus, there are two cases:\begin{enumerate}
			\item  $\mathrm{Ind}\left (B \cap \left ( \bigcup_{k \geq j} A_k \right )\right) = 0$. If this happens, then because the query $B$ is local, it must be the case that $B$ has a cycle in $A_j$. However, when we consider the event $Q_i$, recall that this gives the exact labels corresponding to edges in $A_{i-1}$, including those in $A_j$. Hence, conditioned on $Q_i$, the query $B$ will always have a cycle, and the answer to query $B$ matches $O_i$ conditioned on $Q_i$ with probability $1$.
			\item $\mathrm{Ind}\left (B \cap \left ( \bigcup_{k \geq j} A_k \right )\right) = 1$. In this case, some information about the permutation is leaked, as for every $\ell > j$, $\left (B \cap \left ( \bigcup_{k \geq \ell} A_k \right )\right)$ is an independent set. In particular in the $i$th round, it must be the case that $\left (B \cap \left ( \bigcup_{k \geq i} A_k \right )\right)$ has no cycles.
			However, because the query $B$ was local, it must be the case that for every $\ell > j$, 
			\[
			\left | B \cap \left ( \bigcup_{k \geq \ell} A_k  \right ) \right | \leq \left ( 1 - \frac{100 c \log(L)}{\sqrt{L}\gamma^{\ell}}\right ) \cdot \left | \left ( \bigcup_{k \geq \ell} A_{k} \right ) \right |.
			\]
			In particular, for $\ell = i$, this means that 
			\[
			\left | B \cap \left ( \bigcup_{k \geq i} A_k  \right ) \right | \leq \left ( 1 - \frac{100 c \log(L)}{\sqrt{L}\gamma^{i}}\right ) \cdot \left | \left ( \bigcup_{k \geq i} A_{k} \right ) \right |.
			\] 
			
			Recall now that our goal is to show that conditioned on the size bound on $\left | B \cap \left ( \bigcup_{k \geq i} A_k  \right ) \right |$ above, that there are no cycles in the query $B$. Because the labels over edges in $A_{\geq i}$ are uniformly random, one way to view the above is that $B$ is essentially deleting $\geq \frac{100c\log(L)}{\sqrt{L}\gamma^i}$ random edges from $A_{\geq i}$.
			Because deleting more edges only decreases the probability of a cycle in $B$, we focus our attention on the case when $\left | B \cap \left ( \bigcup_{k \geq i} A_k  \right ) \right | = \left ( 1 - \frac{100c\log(L)}{\sqrt{L}\gamma^{i}}\right ) \cdot \left | \left ( \bigcup_{k \geq i} A_{k} \right ) \right |.$ Thus it remains only to show that if each edge in $\bigcup_{k \geq i} A_k$ is kept with probability $\left ( 1 - \frac{100c\log(L)}{\sqrt{L}\gamma^{i}}\right )$ (equivalently, every edge is deleted with probability $\frac{100c\log(L)}{\sqrt{L}\gamma^{i}}$), then there is a high probability that the resulting graph has no cycles (note that independently deleting edges introduces a multiplicative loss of $\leq$ the number of edges, i.e, $\leq L^2$ in the probability bound we get, as we instead model it with a binomial distribution). However, this follows simply. Each cycle in $A_{\geq i}$ is of length at least $\sqrt{L}\gamma^i$. Thus, the expected number of edges that is deleted is $\geq 100 c \log(L)$. Thus the probability of a cycle surviving the sampling is $\leq 1 / L^{8c}$ (by a Chernoff bound), taking the union bound over all $\leq L^2$ cycles yields that no cycle survives sampling with probability $\geq 1- 1 / L^{6c}$. 
		\end{enumerate}
		
		Thus, we have established that for a fixed query $B$ in the first $i-1$ rounds, conditioned on $B$ being local, the probability that $O_i$ is true for query $B$ conditioned on $Q_i$ is $\geq 1- 1 / L^{6c}$. Because there are at most $L^c \cdot (i-1)$ queries performed in each of the first $i-1$ rounds, we can also take a union bound over each possible query, and see that $\Pr[O_i | Q_i] \geq 1 - 1 / L^{4c}$.

		Now, we will upper bound $\Pr[E | Q_i]$. Let the set of queries that are performed in the $i$th round be denoted by $B_1, \dots B_q$, and we will focus our attention on a single one of these queries. Our goal will be to show that any fixed query is local with high probability, and then we can simply take a union bound over all $q \leq L^c$ queries. We let $m'$ denote the number of edges in $G$ that are in cycles of length $\geq \sqrt{L}\gamma^i$. Thus $B$ is a query that selects $|B|$ out of the $m'$ edges in cycles of length $\geq \gamma^i$. Further, because we are interested in $\Pr[E | Q_i]$, we are exactly told all of the elements of the matroid in cycles of length $\leq \sqrt{L}\gamma^{i-1}$. Next, we let $\beta = \frac{|B|}{m'}$ denote the fraction of remaining edges in $A_{\geq i} $ that are included in the query $B$. Conditioned on $Q_i$, these edges in $A_{\geq i} $ are given uniformly random labels. Thus, instead of considering a query which selects $|B|$ random edges from $A_{\geq i}$ (call this distribution $L_1$), we can instead consider a query which selects each edge in $A_{\geq i}$ independently with probability $\beta$ (call this distribution over queries $L_2$). Note that $\Pr_{L_2}[|B| \text{ edges survive}] \geq \frac{1}{m'+1}$, as $|B|$ is the mean (and therefore mode) of the binomial distribution. Further, conditioned on $|B|$ edges surviving, the distribution induced by $L_2$ is exactly $L_1$. This means that for any event $W$, we have that
		\[
		\Pr_{L_1}[W] \leq (m'+1) \cdot  \Pr_{L_2}[W].
		\]
		
		We let this event $W$ be exactly the event that $B$ is not a local query given $Q_i$. Now, under the distribution $L_2$, we can create a simple case analysis based on $\beta:$
		\begin{enumerate}
			\item If $\beta^{\sqrt{L}\gamma^i} \geq\frac{50c \log(L)}{L}$. If $\beta$ is this large, then we will show that a cycle of length $ \sqrt{L}\gamma^i$ will survive the sampling with high probability. Indeed, the probability a cycle of length $\sqrt{L}\gamma^i$ survives sampling is exactly $\beta^{\sqrt{L}\gamma^i}$ and therefore, 
			\[
			\Pr[\text{any cycle of length }\sqrt{L}\gamma^i \text{ survives sampling}] \geq 1 - (1 - \frac{50 c \log(L)}{L})^L \geq 1 - 1 / L^{8c}.
			\]
			Finally, observe that if the query $B$ in the $i$th round contains a cycle of length $\sqrt{L}\gamma^i$, then the query is necessarily local. 
			\item If $\beta^{\sqrt{L}\gamma^i} \leq \frac{50c \log(L)}{L}$. We first show that under such a sampling rate, there are no cycles of length $\geq \sqrt{L}\gamma^{i+1}$ which survive sampling. This follows simply, as a cycle of length $\geq \sqrt{L}\gamma^{i+1}$ will survive the sampling with probability 
			\[
			\leq \beta^{\sqrt{L}\gamma^{i+1}} = \left ( \beta^{\sqrt{L}\gamma^i} \right )^{\gamma} \leq \left ( \frac{1}{\sqrt{L}}\right )^{\gamma} = \frac{1}{L^{\gamma/2}} \leq \frac{1}{L^{100 c}}.
			\]
			We can then take a simple union bound over all possible $L^2$ cycles, and see that the probability any cycle of length $\geq \sqrt{L}\gamma^{i+1}$ survives sampling is $\leq\frac{1}{L^{98c}}$. Next, we also want to show that not too many edges survive the sampling. For this, we want to show that for all $\ell > i$, 
			\[
			\left | B \cap \left ( \bigcup_{k \geq \ell} A_k  \right ) \right | \leq \left ( 1 - \frac{100c \log(L)}{\sqrt{L}\gamma^{\ell}}\right ) \cdot \left | \left ( \bigcup_{k \geq \ell} A_{k} \right ) \right |.
			\]
			To see why this is the case, let us solve for the value $\beta$ above: indeed if $\beta^{\sqrt{L}\gamma^i} \leq \frac{50c \log(L)}{L}$, then it must be the case that $\beta \leq \left(1 - \frac{\log(L)/10}{\sqrt{L}\gamma^i}\right)$, as if $\beta > \left(1 - \frac{\log(L)/10}{\sqrt{L}\gamma^i}\right)$, we see that 
			\[
			\left(1 - \frac{\log(L)/10}{\sqrt{L}\gamma^i}\right)^{\sqrt{L}\gamma^i} > (1/2)^{\log(L)/10} > 1 / L^{0.1} > \frac{50 c \log(L)}{L},
			\]
			which contradicts our starting assumption. Now, if $\beta \leq \left(1 - \frac{\log(L)/10}{\sqrt{L}\gamma^i}\right)$, then this means that every edge in the graph is deleted with probability $\geq \frac{\log(L)/10}{\sqrt{L}\gamma^i}$. In particular, when we focus on the cycles of length $\ell$ (i.e., $A_{\ell}$), the expected number of deleted edges is at least $|A_{\ell}| \cdot \frac{\log(L)/10}{\sqrt{L}\gamma^i}$. Now, we can calculate the exact number of edges in $A_{\ell}$ to be $L \cdot \sqrt{L} \cdot \gamma^{\ell}$. This means that the expected number of deleted edges is at least 
			\[
			L \cdot \sqrt{L} \cdot \gamma^{\ell} \cdot \frac{\log(L)/10}{\sqrt{L}\gamma^i} \geq \frac{L \log(L) \gamma^{\ell - i}}{10}.
			\]
			In general, for any $\ell > i$, if we look at $\bigcup_{k \geq \ell} A_k$, the expected number of deleted edges is then exactly 
			\[
			\sum_{k \geq \ell} \frac{L \log(L) \gamma^{k - i}}{10},
			\]
			(and the exact number of deleted edges follows a binomial distribution). Recall that our goal was to show that 
			\[
			\left | B \cap \left ( \bigcup_{k \geq \ell} A_k  \right ) \right | \leq \left ( 1 - \frac{100c \log(L)}{\sqrt{L}\gamma^{\ell}}\right ) \cdot \left | \left ( \bigcup_{k \geq \ell} A_{k} \right ) \right |,
			\]
			in particular, this means for any choice of $\ell > i$, we want to show that the number of deleted edges is at least 
			\[
			\frac{100c \log(L)}{\sqrt{L}\gamma^{\ell}} \cdot \left | \left ( \bigcup_{k \geq \ell} A_{k} \right ) \right | = \frac{100c \log(L)}{\sqrt{L}\gamma^{\ell}} \cdot\sum_{k \geq \ell} L \cdot \sqrt{L} \cdot \gamma^{k}.
			\]
			To summarize then, we know that the number of deleted edges follows a binomial distribution with mean $\sum_{k \geq \ell} \frac{L \log(L) \gamma^{k - i}}{10}$, and our goal is to show that this is at least $\frac{100c \log(L)}{\sqrt{L}\gamma^{\ell}} \cdot\sum_{k \geq \ell} L \cdot \sqrt{L} \cdot \gamma^{k}$. In particular, we can observe that $\ell > i$, and so we can re-write the mean as 
			\[
			\geq \sum_{k \geq \ell} \gamma \cdot \frac{L \log(L) \gamma^{k - \ell}}{10} = \frac{L \log(L) \cdot \gamma}{10} \cdot \sum_{k \geq \ell} \gamma^{k- \ell}.
			\]
			We can also re-write our desired number of deletions as 
			\[
			100c \cdot  L \log(L)  \cdot \sum_{k \geq \ell} \gamma^{k - \ell}.
			\]
			Finally, by plugging in our value of $\gamma$, we see that our expected number of deletions is at least 
			\[
			200 c^2 \cdot L \log(L) \cdot \sum_{k \geq \ell} \gamma^{k- \ell} > 2 \cdot (100c \cdot  L \log(L)  \cdot \sum_{k \geq \ell} \gamma^{k - \ell}),
			\]
			and thus is at least double the target number of deletions. Because the mean is at least $L$, we can invoke a Chernoff bound, and obtain that with probability $1 - 2^{-\Omega(L)}$, the number of deletions in $ \left ( \bigcup_{k \geq \ell} A_{k} \right )$ is at least $100c \cdot  L \log(L)  \cdot \sum_{k \geq \ell} \gamma^{k - \ell}$, and thus we can immediately see that with probability at least $1 - 2^{-\Omega(L)}$, for any $\ell > i$, we have \[
			\left | B \cap \left ( \bigcup_{k \geq \ell} A_k  \right ) \right | \leq \left ( 1 - \frac{100c \log(L)}{\sqrt{L}\gamma^{\ell}}\right ) \cdot \left | \left ( \bigcup_{k \geq \ell} A_{k} \right ) \right |,
			\]
			as we desire. Taking a union bound over the $\leq \log(L)$ choices of $\ell$, the statement holds for every $\ell$ with probability $1 - 2^{-\Omega(L)}$.
			
			Finally, to conclude then, we see that in this case, with probability $\geq 1 - 1 / L^{97c}$, our query is local. Indeed, with probability $1 - 1 / L^{98c}$, our query does not yield any cycles of length $\geq \sqrt{L} \gamma^{i+1}$, and with probability $1 - 2^{-\Omega(L)}$, we also satisfy the desired number of missing elements from $\bigcup_{k\geq\ell} A_k$.
		\end{enumerate}
		Thus, we see that in either case, a query $B$ is local with probability $\geq 1 - L^{-8c}$. Taking a union bound over all $\leq L^c$ queries, and accounting for the factor of $m$ (since we assumed that we were sampling edges at rate $\beta$, as opposed to specifying the number of edges exactly), yields that $\Pr[E | Q_i] \leq L^{-5c}$.
		
		To conclude then, we see that 
		\[
		\Pr[E | O_i \cap Q_i] = \frac{\Pr[E \cap O_i | Q_i]}{\Pr[O_i | Q_i]} \leq \frac{\Pr[E| Q_i]}{\Pr[O_i | Q_i]} \leq \frac{L^{-5c}}{1 - \frac{1}{L^{4c}}} \leq 3L^{-4c}.
		\]
	\end{proof}
	
	\begin{claim}\label{clm:manyroundlocal}
		With probability $1 - 3\log(L)L^{-4c}$, all queries in rounds $i \in [\log_{\gamma}(L)/2]$ are local. 
	\end{claim}
	
	\begin{proof}
		This simply follows by taking a union bound over $i \in [\log_{\gamma}(L)/2]$ with \cref{clm:singleRoundLocal}.
	\end{proof}
	
	\begin{proof}[Proof of \cref{thm:mainLB}.]
		By \cref{clm:manyroundlocal}, with probability $1 - 3\log(L)L^{-4c}$, all queries in rounds $i \in [\log_{\gamma}(L)/2]$ are local. In particular, this means that every query $B$ in the first $\log_{\gamma}(L)/2-1$ rounds is local. Any such query $B$ will necessarily not be a complete spanning forest as it either contains a cycle, or is missing too many edges to be a spanning forest (per the definition of \cref{def:local}). Thus, we establish the lower bound of $\Omega(\log(L)) = \Omega(\log(m))$ rounds (using the relationship between $m$ and $L$ as in \cref{def:instance}), as we desire.
	\end{proof}

    Using the same construction, we can extend our lower bound to cographic matroids.

    \begin{corollary}
    Let $A$ be any randomized algorithm which, for cographic matroids on $m$ elements, uses at most $\mathrm{poly}(m)$ queries to an independence oracle per round. Then, in expectation, $A$ must use $\Omega(\log(m))$ adaptive rounds to find a basis of a cographic matroid. 
    \end{corollary}
    \begin{proof}
    Let $G$ be the graphic matroid defined in \cref{def:instance}. By construction, $G$ is the disjoint union of $k=L\cdot \log_\gamma(L)/2$ cycles. We denote $G=\bigoplus _{i=1}^kC_i$, where each $C_i$ is a single cycle, and $\oplus$ denotes the direct sum of matroids on disjoint ground sets. We showed that any algorithm making at most $\mathrm{poly}(m)$ independence queries per round requires $\Omega(\log m)$ adaptive rounds to find a basis of $G$.

    For each cycle $C_i$, suppose it is of length $\ell$. Since $C_i$ is a connected planar graph, its dual matroid is isomorphic to its planar dual graph, which is a multigraph on two vertices connected by $\ell$ parallel edges. We denote this graph by $C_i^*$. In particular, $C_i^*$ is itself a graphic matroid, and the dual of $C_i^*$ satisfies $(C_i^*)^*\cong C_i$. See e.g. Section 2.3 of \cite{Oxl11} for proofs.

    Let $G^*=\bigoplus_{i=1}^k C_i^*$. Then $G^*$ is a graphic matroid consisting of a disjoint union of parallel edges. Its dual matroid $G^*$ is a cographic matroid, and is given by
    \[
    (G^*)^*=\left(\bigoplus_{i=1}^k C_i^*\right)^*\cong\bigoplus_{i=1}^k (C_i^*)^*\cong\bigoplus_{i=1}^k C_i  = G.
    \]
    The first isomorphism uses that fact that the dual of the direct sum of two matroids is (isomorphic to) the direct sum of their duals (see e.g. Proposition 4.2.21 of \cite{Oxl11}).

    Therefore, $G$ is ismorphic to the dual of a graphic matroid, i.e. to a cographic matroid. Since isomorphism preserves independence oracle behavior up to relabeling, it follows that there exists a cographic matroid, for which any randomized algorithm making at most $\mathrm{poly}(m)$ queries to an independence oracle per round requires $\Omega(\log (m))$ adaptive rounds to find a basis.
    \end{proof}
	
	\section{Generalizing to Other Matroids}\label{sec:otherMatroids}
	
	In this section, we show how to generalize our $O(\log(m))$-round algorithm to other classes of matroids. To start, we require the notion of an $\F_2$-representable matroid:
	
	\begin{definition}
		A matroid $M = (E, I)$ is said to be $\F_2$ representable if there is a map from every element $e_i \in E$, to a vector $v_i$ over $\F_2^m$ such that a set $S \subseteq E$ is independent if and only if the corresponding set of vectors $\{v_i: e_i \in S\}$ is linearly independent. 
	\end{definition}
	
	$\F_2$-representability holds true for graphic matroids, and more generally is a subset of the conditions required for \emph{regular} matroids:
	
	\begin{definition}
		A matroid $M = (E, I)$ is said to be \emph{regular} if for every finite field $\F$, $M$ can be represented as a collection of vectors over $\F$.
	\end{definition}
	
	In order for our algorithm to find a basis in $O(\log(m))$ rounds, we require that a matroid is $\F_2$-representable and satisfies a so-called ``smooth counting bound''.
	
	\begin{definition}
		We say that a matroid $M = (E, I)$ satisfies a smooth circuit counting bound if for every $S \subseteq E$, if $M|_S$ has minimum circuit length $\ell$, then for any $\alpha \in \Z^+$, the number of circuits of length $\leq \alpha \cdot \ell$ in $M|_S$ is at most $m^{O(\alpha)}$.
	\end{definition}
	
	With this, in this section we show the following theorem:
	
	\begin{theorem}\label{thm:representableCountingBoundAlg}
		Let $M = (E, I)$ be a matroid representable over $\F_2$ that satisfies a smooth counting bound. Then, there is an $O(\log(m))$-round, $\mathrm{poly}(m)$-query algorithm which finds a basis of $M$ with high probability. 
	\end{theorem}
	
	Several remarks are in order. First, we can immediately apply the above theorem to the class of \emph{cographic matroids}:
	
	\begin{definition}[Cographic Matroids, see, for instance \cite{GV21}.]
		For a graph $G = (V, E)$, its cographic matroid $M$ is the matroid whose independent sets are the subsets of edges whose \emph{complements} contain a spanning forest of $G$. The circuits in $M$ are exactly the cuts of $G$. 
	\end{definition}
	
	Indeed, because the circuits in $M$ are exactly the cuts of $G$, by Karger's \cite{Kar93} famous counting bound, we know that the number of circuits of size $\leq \alpha \ell$ is at most $m^{O(\alpha)}$. In conjunction with the fact that cographic matroids are regular and thus representable over $\F_2$ (see for instance \cite{GV21}), this gives us the following corollary:
	
	\begin{corollary}
		Let $M = (E, I)$ be a cographic matroid. Then, there is an $O(\log(m))$-round, $\mathrm{poly}(m)$-query algorithm which finds a basis of $M$ with high probability. 
	\end{corollary}
	
	For the more general class of regular matroids, while there is no explicit proof that such matroids always obey a smooth counting bound, there is evidence to suggest that such a counting bound is possible. Indeed, the work of \cite{GV21} showed that such matroids obey a ``quadratic'' growth in their circuit sizes. I.e., that the number of circuits of length $\leq \alpha \ell$ (for $\ell$ being the minimum length) is at most $m^{O(\alpha^2)}$.
	
	If this $\alpha^2$ could be improved to $O(\alpha)$, then this would immediately imply an $O(\log(m))$-round $\mathrm{poly}(m)$ query algorithm for finding a basis of $M$. Thus, we leave it as a formal conjecture whether such a counting bound is possible:
	
	\begin{conjecture}\label{conj:regularCountingBound}
		Let $M = (E, I)$ be a regular matroid on $m$ elements with minimum circuit length $\ell$. Then, for any $\alpha \in \Z^+$, $M$ has at most $m^{O(\alpha)}$ circuits of length $\leq \alpha \cdot \ell$.
	\end{conjecture}
	
	As stated above, this conjecture and \cref{thm:representableCountingBoundAlg} together imply the following:
	
	\begin{corollary}
		If \cref{conj:regularCountingBound} is true, then there is an $O(\log(m))$-round, $\mathrm{poly}(m)$-query algorithm for finding a basis of any regular matroid. 
	\end{corollary}

    Likewise, we can pose the same conjecture for the broader class of \emph{max-flow min-cut} matroids, as introduced in the work of \cite{seymour1977matroids}. Such matroids are already known to be binary~\cite{seymour1977matroids}, and, just as in the regular matroid setting, the work of \cite{GV19} showed that the circuits in these matroids obey a counting bound with $O(\alpha^2)$ in the exponent. Improving this bound to $O(\alpha)$ in the exponent would likewise imply an optimal algorithm for basis finding in this class of matroids (and by proxy, the same implication for regular matroids).
	
	With this established, we now proceed to the proof of \cref{thm:representableCountingBoundAlg}.
	
	\subsection{Preliminaries}
	
	Part of the power of looking at regular matroids is due to the fact that they are \emph{representable over} $\F_2$. In general, matroids representable over $\F_2$ satisfy the following property:
	
	\begin{claim}\label{clm:CircuitXOR}
		Let $M = (E, I)$ be a matroid representable over $\F_2$, and let $C_1, C_2 \subseteq E$ denote any two circuits in $M$. Then, $C_1 \oplus C_2$ contains a circuit in $M$.
	\end{claim}
	
	\begin{proof}
		Given $M$, we construct the set of vectors $v_1, \dots v_n$ over $\F_2$ such that $S \subseteq E$ is independent if and only if $\{v_i: i \in S\}$ is linearly independent. By definition, we know that 
		\[
		\sum_{i \in C_1} v_i = 0 = \sum_{i \in C_2} v_i,
		\]
		and so it must also be the case that 
		\[
		\sum_{i \in C_1} v_i + \sum_{i \in C_2} v_i = \sum_{i \in C_1 \oplus C_2} v_i = 0. 
		\]
		Thus, there must be some linear dependence (and thus circuit) that is contained in $C_1 \oplus C_2$.
	\end{proof}
	
	Now, we use the above to establish the following lemma:
	
	\begin{lemma}\label{lem:circuitOverlapGeneral}
		Let $M = (E, I)$ be a matroid representable over $\F_2$ with minimum circuit length $\ell$. Let $C$ be a circuit of length $k \leq 1.01 \ell$ in $E$ and let $C'$ be a circuit of length $k'$ in $E$. Then, $|C' \setminus C| \geq \frac{k'}{4}$.
	\end{lemma}
	
	\begin{proof}
		Note that if $k'$ is sufficiently large relative to $\ell$, the above lemma is trivial. For instance, if $k' \ge 1.5\ell$, then $|C' \setminus C| \ge k' - 1.01\ell \ge k' - (3/4)k' = \frac{k'}{4}$ (using $1.01\ell \le (3/4)k'$ for $k' \ge 1.5\ell$). The interesting case is when $k' \le 1.5\ell$. Here we use \cref{clm:CircuitXOR}: since $C \oplus C'$ must contain a circuit and $\ell$ is the shortest circuit length in $M$, we have $|C \oplus C'| \ge \ell$. We can rewrite 
		\[|C \oplus C'| = |C| + |C'| - 2|C \cap C'| = |C| + |C'| - 2(|C'| - |C' \setminus C|).\] 
		Plugging in $|C \oplus C'| \ge \ell$ and using $|C| \le 1.01\ell$ and $|C'| \ge \ell$, we obtain 
		\[ |C| + |C'| - 2(|C'| - |C' \setminus C|) \ge \ell, \] 
		which simplifies to 
		\[ 2|C' \setminus C| \ge \ell + |C'| - |C| \ge 0.99\ell. \] 
		Thus $|C' \setminus C| \ge 0.495\ell$, which in this case is at least $k'/4$. 
	\end{proof}
	
	With these structural claims established, we can now mirror the algorithm for finding bases of graphic matroids. We re-present this algorithm and its analysis in the following subsections.
	
	\subsection{Detecting a Single Circuit}
	
	As in the graphic matroid case, our algorithm proceeds by removing circuits in an iterative manner, gradually eliminating circuits of increasing lengths. Suppose in some iteration of the algorithm we have the promise that there are no circuits of length $\leq \ell$ in the matroid. Then our goal for the iteration is to (1) eliminate all circuits of length $\leq 1.01 \ell$ and (2) to do this \emph{without} altering the rank of the matroid. We will accomplish this task by repeatedly sampling the elements in the matroid to create the following {\em good event}: that there is a unique circuit that survives among the sampled elements and moreover, that its length is $\leq 1.01 \ell$.  
	
	Conditioned on this good event, we must then identify exactly the elements that are participating in this unique surviving circuit, and then repeat this process many times until we have enumerated all circuits of length $\leq 1.01 \ell$ in the matroid. Note that because of our assumption that the matroid satisfies a smooth counting bound, as a sanity check we can see that the number of potential circuits we must recover is bounded by some polynomial in $m$ (although there is no guarantee that these circuits are easy to find). 
	
	As a first step towards identifying these circuits, we present a simple algorithm for detecting whether or not there is a single circuit in a matroid (and if there is a single circuit, the algorithm returns exactly the elements in the circuit):
	
	\begin{algorithm}[H]
		\caption{DetectSingleCircuit($E'$)}\label{alg:detectSingleCircuit}
		Initialize the set of critical elements $S = \emptyset$.\\
		\If{\text{Query }$\mathrm{Ind}(E') = 1$}{
			\Return{$\perp$, No circuits.}
		}
		\For{$e \in E'$}{
			\If{\text{Query }$\mathrm{Ind}(E' \setminus \{e\}) = 1$}{
				$S \leftarrow S \cup \{ e \}$.
			}
		}
		\If{$S = \emptyset$}{
			\Return{$\perp$, $\geq 2$ circuits.}
		}
		\Return{$S$.}
	\end{algorithm}
	
	We summarize the performance of the algorithm in several claims:
	
	\begin{claim}\label{clm:detectSingleCircuit1}
		Let $E'$ be a set of elements which has no circuits, then \cref{alg:detectSingleCircuit} correctly returns that there are no circuits.
	\end{claim}
	
	\begin{proof}
		The independence query to $E'$ will be $1$ if and only if there are no circuits.
	\end{proof}
	
	\begin{claim}\label{clm:detectSingleCircuit2}
		Suppose $E'$ has $\geq 2$ circuits, then \cref{alg:detectSingleCircuit} returns that there are $\geq 2$ circuits.
	\end{claim}
	
	\begin{proof}
		To prove this, we use a well-known fact: namely that if a matroid $M$ has at least $2$ distinct circuits, then there is no single element whose removal kills both circuits. To see why, let us denote two circuits by $C_1, C_2$. If an element $e$ is in both $C_1$ and $C_2$, then $C_1 \cup C_2 \setminus \{e\}$ also contains a circuit (importantly, which $e$ is not in). Thus, there must be some circuit in the matroid which does not include the element $e$, so $e$'s removal does not make the remaining elements independent. Otherwise, if $e$ is not in both of $C_1, C_2$, then one of $C_1, C_2$ will also remain intact after $e$'s removal. This yields the claim.
	\end{proof}
	
	\begin{claim}\label{clm:detectSingleCircuit3}
		Suppose $E'$ has exactly $1$ circuit, then \cref{alg:detectSingleCircuit} returns exactly the constituent elements of this circuit. 
	\end{claim}
	
	\begin{proof}
		Let the circuit be denoted by $C$. Observe that removing any element $e \in C$ will remove the dependence in the circuit (now the set of elements in independent by definition), and hence the independence queries will now return $1$. Likewise, deleting any element $e \notin C$ will not remove the circuit, and the circuit will still be present, so the independence queries will return $0$. 
	\end{proof}
	
	Finally, we also observe that the above algorithm can be implemented in $1$ round of adaptivity:
	
	\begin{claim}\label{clm:detectSingleCircuit4}
		\cref{alg:detectSingleCircuit} can be implemented in $1$ round of adaptivity and makes $|E'| + 1$ queries to $\mathrm{Ind}$.
	\end{claim}
	
	\begin{proof}
		Notice that the algorithm queries $\mathrm{Ind}(E')$, and $\mathrm{Ind}(E' \setminus \{e\}): \forall e \in E'$. All these queries are made without reference to the results from previous queries.
	\end{proof}
	
	Thus, we get the following lemma to summarize \cref{alg:detectSingleCircuit}:
	
	\begin{lemma}\label{lem:detectSingleCircuit}
		For a set of elements $E'$, \cref{alg:detectSingleCircuit} makes $|E'| +1$ queries to $\mathrm{Ind}$ in only a single round, and returns $\perp$ if $E'$ has no circuits or $\geq 2$ circuits, and otherwise returns a set $S \subseteq E'$ which is exactly the elements involved in the single circuit in $E'$.
	\end{lemma}
	
	\begin{proof}
		This follows from \cref{clm:detectSingleCircuit1}, \cref{clm:detectSingleCircuit2}, \cref{clm:detectSingleCircuit3}, and \cref{clm:detectSingleCircuit4}.
	\end{proof}
	
	Thus, we have a simple algorithm for identifying when there is a single circuit in a matroid. In the coming sections, we will show how to use this procedure to identify and remove all of the short circuits in a given matroid, provided the matroid satisfies certain structural properties. 
	
	\subsection{Sampling}
	
	Motivated by the previous subsection, our goal will now be to sub-sample the matroid at a specific rate such that only \emph{one} circuit will survive the sampling process with high probability. When only one circuit survives, we can then identify this circuit exactly by \cref{alg:detectSingleCircuit}. The sampling algorithm is provided below, which takes in a number of elements $m$, the set of elements in the matroid $E$, as well as a parameter $\ell$ corresponding to the minimum circuit length in $E$.
	
	\begin{algorithm}[H]
		\caption{RecoverCircuitSuperset$(E, m, \ell)$}\label{alg:recoverCircuit}
		Let $p$ be such that $p^{\ell} = \frac{1}{m^{100 \kappa}}$, for $\kappa$ a sufficiently large constant. \\
		\For{$i \in [m^{102\kappa}]$, in parallel }{
			Let $E^{(i)}$ be the result of independently keeping each $e \in E$ with probability $p$.
		}
		\Return{$\{E^{(i)}: i \in [m^{102\kappa}]\}$}
	\end{algorithm}

	To understand the sampling procedure, we focus on a single circuit $C$ in $E$ of length $k$, for $k \in (\ell, 1.01 \ell]$, and start by showing the following:
	
	\begin{claim}\label{clm:uniqueCircuitProb}
		Let $E$ be a set of $m$ elements from an $\F_2$-representable matroid $M$ satisfying a smooth counting bound. Suppose further that there is no circuit of length $\leq \ell$ and let $C$ be a circuit in $E$ of length $k$, for $k \in (\ell, 1.01 \ell]$. Let $E'$ be the result of independently keeping each element in $E$ with probability $p$, where $p^{\ell} = \frac{1}{m^{100\kappa}}$ (for $\kappa$ a sufficiently large constant). Then, $C$ is the unique circuit in $E'$ with probability $\frac{1}{2\cdot m^{101\kappa}}$.
	\end{claim}
	
	\begin{proof}
		First, we calculate the probability that $C$ survives the sampling procedure. This is straightforward, as $C$ has $\leq 1.01 \ell$ elements. So, after sampling at rate $p$, the probability all $\leq 1.01 \ell$ elements survive is:
		\[
		\Pr[C \text{ survives sampling}] \geq p^{1.01 \ell} = \left ( p^{\ell} \right )^{1.01} = \left ( \frac{1}{m^{100\kappa}} \right )^{1.01} = \frac{1}{m^{101\kappa}}.
		\]
		
		Thus, it remains only to bound the probability that \emph{some other circuit} $C'$ also survives the sampling at rate $p$, conditioned on $C$ surviving the sampling. 
		
		Indeed, to bound this, let us suppose that $C'$ is of length $k'$. By \cref{lem:circuitOverlapGeneral}, we know that $|C' \setminus C| \geq \frac{k'}{4}$. Thus,
		\[
		\Pr[C' \text{ survives sampling}  | C \text{ survives sampling}] \leq p^{k'/4},
		\]
		as there will be at least $k'/4$ elements in $C'$ which are not in $C$ (and thus these elements surviving the sampling process is independent of $C$ surviving). 
		
		Next, it remains to take a union bound over all possible circuits $C'$. For this, let $\alpha$ be the power of $2$ such that $k' \in [\alpha \ell, 2\alpha \ell]$, and then we use our assumption that $M$ satisfies a smooth counting bound. Specifically, for circuits $C'$ of length $[\alpha \ell, 2\alpha \ell]$, we know that there are at most $m^{O(\alpha)}$ such circuits. So, for a fixed $\alpha$, this means we get the following bound:
		\[
		\Pr[\exists C'\neq C \text{ of length }[\alpha \ell, 2\alpha \ell] \text{ that survives sampling} | C \text{ survives sampling}] 
		\]
		\[
		\leq \sum_{C'\neq C: \text{ circuit of length }[\alpha \ell, 2\alpha \ell]} \Pr[C' \text{ survives sampling}  | C \text{ survives sampling}]
		\]
		\[
		\leq \sum_{C': \text{ circuit of length }[\alpha \ell, 2\alpha \ell]} p^{\alpha \ell/4} \leq p^{\alpha \ell/4} \cdot m^{O(\alpha)} = \left ( \frac{1}{m^{100\kappa}} \right )^{\frac{\alpha}{4}} \cdot m^{O(\alpha)} \leq \left (\frac{1}{m^{23\kappa}} \right)^{\alpha},
		\]
		by choosing $\kappa$ to be a sufficiently large constant.
		
		To conclude, we can then take a union bound over $\alpha \in \{1, 2, 4, 8, \dots m \}$. Thus, we see that:
		\[
		\Pr[\exists C'\neq C\text{ that survives sampling} | C \text{ survives sampling}] 
		\]
		\[
		\leq \sum_{\alpha \in \{1, 2, 4, 8, \dots m \}} \Pr[\exists C'\neq C \text{ of length }[\alpha \ell,2\alpha \ell] \text{ that survives sampling} | C \text{ survives sampling}] 
		\]
		\[
		\leq \sum_{\alpha \in \{1, 2, 4, 8, \dots m \}} \left (\frac{1}{m^{23\kappa}} \right)^{\alpha} \leq \sum_{\alpha = 1}^{\infty} \left (\frac{1}{m^{23\kappa}} \right)^{\alpha} 
		\]
		\[
		\leq \frac{2}{m^{23\kappa}}
		\]
		where the final inequality follows because the expression is a geometric series with ratio $< 1/2$.
		
		Finally, recall that our goal was to show that $C$ is the unique circuit which survives sampling with non-negligible probability. For this, observe that 
		\[
		\Pr[C \text{ survives sampling}] = \Pr[C \text{ uniquely survives sampling}] + \Pr[\exists C'\neq C: C \wedge C' \text{ survive sampling}].
		\]
		Now, this second term we can bound by our above work. I.e.,
		\[
		\Pr[\exists C'\neq C: C \wedge C' \text{ survive sampling}]
		\]
		\[
		= \Pr[\exists C'\neq C\text{ that survives sampling} | C \text{ survives sampling}]\cdot\Pr[C \text{ survives sampling}]
		\]
		\[
		\leq \frac{2}{m^{23\kappa}} \cdot \Pr[C \text{ survives sampling}].
		\]
		Thus, we see that 
		\[
		\Pr[C \text{ uniquely survives sampling}] \geq \Pr[C \text{ survives sampling}] \cdot \left ( 1 - \frac{2}{m^{23\kappa}}\right ) \geq \frac{1}{2\cdot m^{101\kappa}},
		\]
		as we desire.
	\end{proof}
	
	Using the above claim, we now show that (with high probability) every circuit $C$ of length $(\ell, 1.01 \ell]$ is the unique surviving circuit for some $E^{(i)}$ produced by \cref{alg:recoverCircuit}.
	
	\begin{lemma}\label{lem:uniqueCircuitSurvive}
		Let $E$ be a set of $m$ elements in an $\F_2$-representable matroid $M$ satisfying a smooth counting bound such that there is no circuit of length $\leq \ell$. Then, with probability $1 - 2^{-\Omega(m)}$, for every circuit $C$ in $E$ of length $(\ell, 1.01 \ell]$, there is an index $i \in [m^{102\kappa}]$ such that $C$ is the unique circuit in $E^{(i)}$.
	\end{lemma}
	
	\begin{proof}
		Fix any circuit $C$ in $E$ of length $(\ell, 1.01 \ell]$. By \cref{clm:uniqueCircuitProb}, we know that over the randomness of the sampling procedure, $C$ will be the unique circuit present in $E^{(i)}$ with probability $\frac{1}{2m^{101\kappa}}$. Thus, by repeating this procedure $m^{102\kappa}$ times, we know that there is at least one index $i$ for which $C$ is the unique circuit with probability $1 - 2^{-\Omega(m)}$.
		
		Now, because there are at most $m^{O(1)}$ circuits of length $(\ell, 1.01\ell]$ in a matroid with no circuits of length $\leq \ell$, we can take a union bound over all these circuits. This means that with probability $1 - 2^{-\Omega(m)}$, for every circuit $C$ in $E$ of length $(\ell, 1.01 \ell]$, there is an index $i \in [m^{102\kappa}]$ such that $C$ is the unique circuit in $E^{(i)}$.
	\end{proof}

	\subsection{Final Circuit Recovery Algorithm}
	
	Now, it remains only to piece together these algorithms. We present this below as an algorithm:
	
	\begin{algorithm}[H]
		\caption{TotalCircuitRecovery$(E, m, \ell)$}\label{alg:finalCircuit}
		Let $p$ be such that $p^{\ell} = \frac{1}{m^{100\kappa}}$. \\
		$\mathrm{Circuits}= \{\}$.\\
		\For{$i \in [m^{102\kappa}]$ in parallel}{
			Let $E^{(i)}$ be the result of independently keeping each $e \in E$ with probability $p$. \\
			Let $S = \mathrm{DetectSingleCircuit}(E^{(i)})$. \\
			\If{$S \neq \perp$}{
				$\mathrm{Circuits} \leftarrow \mathrm{Circuits} \cup \{S\}$.
			}
		}
		\Return{$\mathrm{Circuits}$.}
	\end{algorithm}
	
	We can combine our results from the previous subsections to understand what \cref{alg:finalCircuit} achieves:
	
	\begin{lemma}\label{lem:recoverShortCircuits}
		Let $E$ be a set of $m$ elements in an $\F_2$-representable matroid $M$ satisfying a smooth counting bound, such that there is no circuit of length $\leq \ell$ in $E$. Then, with probability $1 - 2^{-\Omega(m)}$, the output of \cref{alg:finalCircuit} is a set of circuits which contains every circuit $C$ in $E$ of length $(\ell, 1.01 \ell]$.
	\end{lemma}
	
	\begin{proof}
		First, recall that by \cref{lem:detectSingleCircuit}, the algorithm DetectSingleCircuit does not return $\perp$ if and only if there is a single circuit in the input matroid. By this same lemma, when the output is not $\perp$, the algorithm recovers exactly the constituent elements of the circuit. Hence the output of \cref{alg:finalCircuit} is necessarily a set of circuits (as the circuits in the subsampled matroids will also be circuits in the original matroid). It remains only to show that \emph{every} circuit of length $(\ell, 1.01 \ell]$ is included in the output. This follows exactly from \cref{lem:uniqueCircuitSurvive}. Indeed, every circuit of length $(\ell, 1.01 \ell]$ will be the unique circuit surviving the sampling procedure in some round with probability $1 - 2^{-\Omega(m)}$, and thus will be included in the output as well. This yields the lemma at hand. 
	\end{proof}
	
	Likewise, we can observe that \cref{alg:finalCircuit} is implementable in parallel without any adaptivity:
	
	\begin{claim}
		On input $E$, with $m$ elements, and parameter $\ell$, \cref{alg:finalCircuit} is implementable with $\mathrm{poly}(m)$ queries to the independence oracle in a single round.
	\end{claim}
	
	\begin{proof}
		By \cref{lem:detectSingleCircuit}, each invocation of DetectSingleCircuit requires $O(m)$ queries to the independence oracle. Each of the sub-sampled matroids is checked in parallel, and thus the total number of queries is $O(m \cdot m^{102\kappa}) = \mathrm{poly}(m)$, without any adaptivity. 
	\end{proof}
	
	\subsection{Circuit Removal}
	
	After recovering all of the circuits of length $(\ell, 1.01\ell]$, we must find a way to delete these circuits without altering the rank of the matroid. To do this, we take advantage of a basic operation from the work of \cite{KUW85}. Although this statement is included in \cite{KUW85}, there was no proof provided there, so we re-prove the result here.
	
	\begin{lemma}\label{lem:deleteCircuits}
		Let $E$ be a set of elements with some fixed ordering of the elements $e_1, \dots e_m$, and let $\mathrm{Circuits}$ be an arbitrary subset of the circuits in $E$. For each circuit $C \in \mathrm{Circuits}$, let $C = (e_{i_{C, 1}}, \dots e_{i_{C, |C|}})$ denote the ordered set of the elements that are in the circuit $C$. Let $E'$ be the result of simultaneously deleting from $E$ the element with the \emph{largest} index from every circuit in $\mathrm{Circuits}$. Then,
		\begin{enumerate}
			\item Every circuit in $\mathrm{Circuits}$ has at least one element removed.
			\item The rank of $E'$ is the same as the rank of $E$. 
		\end{enumerate}
	\end{lemma}
	
	\begin{proof}
		The first item is essentially trivial. Every circuit in $\mathrm{Circuits}$ has some constituent element deleted.
		
		The second item is less immediate. To see why the rank of $E$ and $E'$ is the same, let us consider adding the elements in $E \setminus E'$ back in to $E'$, in increasing order (i.e., starting with the element which has the smallest label). Let us denote these elements by $e_1^*, \dots e_{g}^{*}$. We claim that the rank of $E' \cup \{e_1^*, \dots e_{i}^* \}$ is the same as the rank of $E' \cup \{e_1^*, \dots e_{i+1}^* \}$. For the base case, we show that the rank of $E'$ is the same as the rank of $E' \cup \{e^*_1 \}$. This is because $e^*_1$ was the smallest labeled element which was removed. Because it was removed though, this means that there is some circuit $C_1$ for which $e^*_1$ has the largest label, and hence all other elements in $C_1 \setminus \{ e^*_1\}$ are still present in $E'$. But, adding $e^*_1$ then does not alter the rank, as $e^*_1$ is dependent with respect to $C_1 \setminus \{ e^*_1\}$.
		
		Now, we show the general case. I.e., that the rank of $E' \cup \{e_1^*, \dots e_{i}^* \}$ is the same as the rank of $E' \cup \{e_1^*, \dots e_{i+1}^* \}$. For this, observe that because the $e_j^*$'s are ordered in terms of their labels, $E' \cup \{e_1^*, \dots e_{i}^* \}$ contains every element whose label is smaller than $e_{i+1}^*$. In particular, because $e_{i+1}^*$ was removed, this means that $e_{i+1}^*$ was the largest labeled element in some circuit $C_{i+1}$. This means that every element in $C_{i+1} \setminus \{ e_{i+1}^*\}$ is still in $E' \cup \{e_1^*, \dots e_{i}^* \}$. But, adding $e^*_{i+1}$ then does not alter the rank, as $e^*_{i+1}$ is dependent with respect to $C_{i+1} \setminus\{ e_{i+1}^*\}$.
		
		To conclude, we simply observe that by induction, this means that the rank of $E'$ is the same as the rank of  $E' \cup \{e_1^*, \dots, e_g^* \}=E$.
	\end{proof}
	
	\subsection{Finding the Basis}
	
	With all of our building blocks now established, we present our final algorithm for finding a basis below.
	
	\begin{algorithm}[H]
		\caption{FindSpanningForest$(E, m)$}\label{alg:findBasis}
		Let $e_1, \dots e_m$ be a fixed ordering of the elements of $E$. \\
		$\ell = 1$. \\
		\While{$|E| \geq \ell$}{
			$\mathrm{Circuits} = \mathrm{TotalCircuitRecovery}(E, |E|, \ell)$. \\
			\For{$C \in \mathrm{Circuits}$ in parallel}{
				Let $e^*$ be the element in $C$ with the largest index with respect to the ordering. \\
				$E \leftarrow E \setminus \{ e^*\}$.
			}
			
			$\ell \leftarrow 1.01 \cdot \ell$.
		}
		\Return{$E$.}
	\end{algorithm}
	
	We now present a sequence of claims analyzing the above algorithm.
	
	\begin{claim}\label{clm:singleIterationCircuit}
		Let $E$ be a set of $m$ elements in an $\F_2$-representable matroid $M$ satisfying a smooth counting bound, with no circuits of length $\leq \ell$. For each iteration of the while loop (Line 3) in \cref{alg:findBasis}, the algorithm removes all circuits of length $\leq 1.01 \ell$ without altering the rank of $E$, with probability $1 - 2^{-\Omega(m)}$.
	\end{claim}
	
	\begin{proof}
		This set of circuits recovered includes all circuits of length $\leq 1.01 \ell$ with probability $1 - 2^{-\Omega(m)}$ by \cref{lem:recoverShortCircuits}. These circuits are then deleted without altering the rank as per \cref{lem:deleteCircuits}.
	\end{proof}
	
	Next, we also bound the number of iterations of the while loop:
	
	\begin{claim}\label{clm:numberRoundsCircuits}
		After invoking \cref{alg:findBasis} on a set of $m$ elements in an $\F_2$-representable matroid $M$ satisfying a smooth counting bound, the while loop in line 14 runs for at most $O(\log(m))$ iterations, with probability $1 - 1 / \mathrm{poly}(m)$, and returns a set of elements with no circuits of length $\leq m+1$.
	\end{claim}
	
	\begin{proof}
		Note that after each iteration of the while loop in line 14, the minimum circuit length increases from $\ell$ to $1.01 \ell$ (as per \cref{clm:numberRoundsCircuits}). In particular, after $O(\log(m))$ iterations, the minimum circuit length will be $\geq m + 1$, at which point there are no more circuits in the entire matroid, and we have reached the stopping condition, as $|E| < \ell$ at this point.

		Note that the probability bound holds because if the number of elements ever drops to below $\log^2(m)$, then we can simply \emph{directly solve} for a basis of the remaining elements using \cite{KUW85} (deterministically). Otherwise, the number of remaining elements is always $\geq \log^2(m)$, and so every iteration of the above procedure succeeds with probability $1 - 1 / 2^{\Omega(\log^2(m))}$. Taking a union bound over all $O(\log(m))$ rounds then yields the claim. 
	\end{proof}
	
	We conclude with our primary theorem:
	
	\begin{proof}[Proof of \cref{thm:representableCountingBoundAlg}]
		To see the bound on the number of queries, and rounds of adaptivity, recall that each invocation of $\mathrm{DetectSingleCircuit}$ requires $O(m)$ queries to the oracle by \cref{lem:detectSingleCircuit}. In each round, there are $\mathrm{poly}(m)$ invocations made to \cref{lem:detectSingleCircuit} (in parallel), and thus the total number of queries per round is $\mathrm{poly}(m)$. Likewise, by \cref{clm:numberRoundsCircuits}, the number of rounds required is $O(\log(m))$, and returns a basis of $E$ with probability $1 - 1 / \mathrm{poly}(m)$. This yields the theorem. Thus \cref{alg:finalCircuit} is indeed an algorithm satisfying the conditions of \cref{thm:representableCountingBoundAlg}.
	\end{proof}
	
	\section{Conclusions}
	
	We have resolved a longstanding open question regarding the parallel complexity of computing bases in graphic matroids. Specifically, we presented a deterministic parallel algorithm that, given access to an independence oracle, computes a spanning forest in $O(\log m)$ adaptive rounds using only $\mathrm{poly}(m)$ non-adaptive queries per round. This matches our lower bound, which shows that any (even randomized) algorithm using $\mathrm{poly}(m)$ queries per round must require $\Omega(\log m)$ adaptive rounds in expectation. Together, these results provide a tight and complete characterization of the round-query tradeoff for graphic matroids.
	
	Our approach introduces new structural insights into the behavior of cycles in graphs under random sampling, as well as techniques for efficiently enumerating and eliminating short cycles while preserving graph connectivity. These tools may find applications in development of parallel algorithms for other graph problems.

	\bibliographystyle{alpha}
	\bibliography{ref}
    
\end{document}